\documentclass[aps,prd,twocolumn,superscriptaddress,preprintnumbers,nofootinbib,10pt]{revtex4-1}
\pdfoutput=1

\usepackage{graphicx}
\usepackage{mathrsfs}
\usepackage{amsmath,amstext,amsfonts,amssymb}
\usepackage{color}

\definecolor{darkblue}{rgb}{0,0,0.5}
\definecolor{blue}{rgb}{0,.3,0.9}
\usepackage[colorlinks,linkcolor=darkblue,citecolor=darkblue,urlcolor=darkblue]{hyperref}

\allowdisplaybreaks

\newcommand{\beq}{\begin{equation}}
\newcommand{\eeq}{\end{equation}}
\newcommand{\bea}{\begin{eqnarray}}
\newcommand{\eea}{\end{eqnarray}}

\newcommand{\GeV}{\text{GeV}}

\newcommand{\Fermi}{{\it Fermi}\ }
\def\missET {{\not\!\! E_T}}

\begin{document}

\preprint{EFI-15-9}

\title{Pseudoscalar Portal Dark Matter}


\author{Asher~Berlin}
\affiliation{Department of Physics, Enrico Fermi Institute, University of Chicago, Chicago, IL 60637}
\affiliation{Kavli Institute for Cosmological Physics, University of Chicago, Chicago, IL 60637}

\author{Stefania~Gori}
\affiliation{Perimeter Institute for Theoretical Physics,\\31 Caroline St. N, Waterloo, Ontario, Canada N2L 2Y5.}

\author{Tongyan~Lin}
\affiliation{Department of Physics, Enrico Fermi Institute, University of Chicago, Chicago, IL 60637}
\affiliation{Kavli Institute for Cosmological Physics, University of Chicago, Chicago, IL 60637}

\author{Lian-Tao~Wang}
\affiliation{Department of Physics, Enrico Fermi Institute, University of Chicago, Chicago, IL 60637}
\affiliation{Kavli Institute for Cosmological Physics, University of Chicago, Chicago, IL 60637}

\date{\today}

\begin{abstract}
A fermion dark matter candidate with a relic abundance set by annihilation through a pseudoscalar can evade constraints from direct detection experiments. We present simplified models that realize this fact by coupling a fermion dark sector to a two-Higgs doublet model.  These models are generalizations of mixed bino-Higgsino dark matter in the MSSM, with more freedom in the couplings and scalar spectra. Annihilation near a pseudoscalar resonance allows a significant amount of parameter space for thermal relic dark matter compared to singlet-doublet dark matter, in which the fermions couple only to the SM Higgs doublet. In a general two-Higgs doublet model, there is also freedom for the pseudoscalar to be relatively light and it is possible to obtain thermal relic dark matter candidates even below 100 GeV.  In particular, we find ample room to obtain dark matter with mass around 50 GeV and fitting the Galactic Center excess in gamma-rays. This region of parameter space can be probed by LHC searches for heavy pseudoscalars or electroweakinos, and possibly by other new collider signals.
\end{abstract}

\maketitle

\section{Introduction}

Is weak-scale dark matter (DM) still a viable scenario? At face value, a DM candidate of mass 10-1000 GeV with weak-scale interactions can easily have a thermal relic abundance matching the observed value of $\Omega_{cdm} h^2 \approx 0.1199 \pm 0.0027$ \cite{Ade:2013zuv}. However, the annihilation process is generically related by crossing symmetry to interactions in direct detection experiments and at colliders, both of which are becoming increasingly restrictive for this range of DM masses. To satisfy these constraints requires more ingredients or tunings in many existing models.

One simple and attractive possibility is that the DM interaction with Standard Model (SM) particles is suppressed in the non-relativistic limit in the $t$-channel (direct detection) while preserving a weak-scale cross section in the $s$-channel (for a thermal relic). This requirement is satisfied by a pseudoscalar mediator coupling to fermion DM and to SM fermions:
\begin{equation}
	{\cal  L} \supset  y_\chi A \bar \chi  i \gamma^5 \chi  + \lambda_f A \bar f  i \gamma^5 f ~.
	\label{eq:Lsimple}
\end{equation}
Because the pseudoscalar interaction breaks chiral symmetry, the coupling of $A$ with SM fermions is generically proportional to the SM Yukawa couplings $y_f$ ($\lambda_f\propto y_f$).  Therefore, since the pseudoscalar has larger couplings to third-generation quarks and leptons, collider constraints are typically weaker. Integrating out the pseudoscalar gives the dimension six operators
\begin{equation}
	\frac{y_\chi \lambda_f}{m_A^2} (\bar \chi \gamma^5 \chi)  ( \bar f \gamma^5 f)~.
\end{equation}
These contact operators have been considered in Refs.~\cite{Lin:2013sca,Bai:2010hh,Goodman:2010ku,Goodman:2010yf,Rajaraman:2011wf,Haisch:2012kf}, which motivated collider signals with MET and a single jet or $b$-jet as a new search channel for DM.

In order to provide a concrete but simple realization of this interaction, the approach taken in this paper is to build a simplified model of DM coupled to a new pseudoscalar mediator. The philosophy is to add a minimal set of new matter fields with renormalizable and gauge-invariant couplings \cite{Abdallah:2014hon}. Our models also provide a UV-completion (at least at scales relevant to the LHC) of the type of contact interactions discussed above.

Since the SM does not contain a fundamental pseudoscalar, we focus our study on two-Higgs doublet models (2HDM). We note that the simplest cases of Higgs-portal models connecting DM to the SM through Higgs mediation are highly constrained \cite{LopezHonorez:2012kv,Greljo:2013wja,Fedderke:2014wda}. In particular, existing models have studied singlet-doublet fermion sectors coupled to the SM Higgs \cite{Enberg:2007rp,Cohen:2011ec,Cheung:2013dua}, while our work generalizes this to 2HDMs. Within the 2HDM framework, the pseudoscalar interaction with DM carries unavoidable interactions of DM to the CP-even scalars as well. However, compared to the usual Higgs portal models, in a general 2HDM, it is possible to obtain parametrically larger couplings of DM to the pseudoscalar (or heavy Higgs) than to the light Higgs, in this way alleviating the various constraints.

The possibility of resonant annihilation through the pseudoscalar also permits a more sizable parameter space for DM to be a thermal relic.  As we will show, much of the open parameter space is in the region where $m_\chi \sim m_A/2$. This is the analog of the so-called ``$A$-funnel" region in the Minimal Supersymmetric Standard Model (MSSM). Still, it is important to note that, even if the DM mass is not tuned to be very close to $m_A/2$, we can obtain a thermal DM and satisfy the constraints from direct detection and from the LHC.

Another motivation to consider models with DM mainly annihilating through the $A$-funnel comes from the intriguing results for DM annihilation in the gamma-ray sky~\cite{Goodenough:2009gk,Hooper:2010mq,Hooper:2011ti,Abazajian:2012pn,Gordon:2013vta,Abazajian:2014fta,Daylan:2014rsa}. Studies of the \Fermi gamma-ray data show an excess diffuse component around the Galactic Center (GC). Although these diffuse gamma-rays may have an astrophysical origin, it is interesting to note that their spectrum and morphology is consistent with expectations for gamma-rays from DM annihilation. The spectrum peaks at around 2-3 GeV, which suggests a DM candidate with mass smaller than around $100$ GeV, and the signal follows an NFW profile. Many theoretical studies have explored simplified scenarios~\cite{Berlin:2014tja,Alves:2014yha,Balazs:2014jla,Berlin:2014pya,Martin:2014sxa}, as well as more UV-complete models, for the Galactic Center Excess (GCE). 
In particular, light DM with mass around $\sim 50$ GeV annihilating mainly to bottom quarks and tau leptons has identified as giving a good fit to the data. However, a more recent  \Fermi analysis of the GC region~\cite{NewFermi} indicates the spectrum of the diffuse component is harder than previously thought, allowing for reasonable fits with heavier DM candidates annihilating to di-boson final states~\cite{Agrawal:2014oha,Calore:2014xka,Calore:2014nla}.

A number of previous works have identified the pseudoscalar case as a promising candidate to fit the GCE~\cite{Izaguirre:2014vva,Cheung:2014lqa,Cahill-Rowley:2014ora,Ipek:2014gua,Huang:2014cla,Hektor:2014kga,Arina:2014yna,Dolan:2014ska,Abdullah:2014lla,Boehm:2014hva,Ghorbani:2014qpa}. In a number of these models, in order to satisfy various experimental constraints, a new light pseudoscalar which mixes with the pseudoscalar in the 2HDM is introduced, while the new scalars in the 2HDM itself are relatively decoupled~\cite{Cheung:2014lqa,Cahill-Rowley:2014ora,Ipek:2014gua,Cao:2014efa}. The new feature of our model is that we rely purely on the pseudoscalar of the 2HDM, but consider more general 2HDMs and dark sector spectra.

The paper is organized as follows. In Section~\ref{2HDM} we review aspects of general 2HDMs and we fix our notation. Furthermore, we motivate the scalar spectra considered in the following sections. Section~\ref{sec:collider} discusses the model-independent collider constraints on the simplified model, covering heavy Higgs searches and searches with MET. We then turn to specific models, beginning with a brief discussion on extended Higgs portal models  for both scalar and fermion DM  in Section~\ref{sec:scalarDM}.  Our main analysis is contained in Section~\ref{themodel} for fermion DM coupled to a 2HDM. We consider a range of constraints on the model, and present viable parameter space both for heavier DM ($\sim$150 GeV) in Section~\ref{discussion} and specifically for the GCE in Section~\ref{GCE}. We reserve Section~\ref{sec:conc} for our conclusions. In the appendices, we elaborate on 2HDM benchmarks for the GCE (Appendix~\ref{app:Higgsspectrum}) and give analytic formulae for Higgs couplings to DM in our model (Appendix~\ref{app:higgscouplings}).

\section{Two Higgs Doublets}
\label{2HDM}

The most general renormalizable Higgs potential of a
2HDM can be written as 
\begin{eqnarray}\label{eq:Higgs_potential}
V &=& m_d^2\Phi_d^\dagger \Phi_d + m_u^2 \Phi_u^\dagger \Phi_u  \\
&& + \frac{\lambda_1}{2} (\Phi_d^\dagger \Phi_d)^2 + \frac{\lambda_2}{2} (\Phi_u^\dagger \Phi_u)^2 \nonumber \\
&& + \lambda_3 (\Phi_d^\dagger \Phi_d)(\Phi_u^\dagger \Phi_u) + \lambda_4 (\Phi_d^\dagger \Phi_u)(\Phi_u^\dagger \Phi_d) \nonumber \\
&& + \Big[ -B\mu (\Phi_d^\dagger \Phi_u) + \frac{\lambda_5}{2} (\Phi_d^\dagger \Phi_u)^2 \nonumber \\
&& + \lambda_6 (\Phi_d^\dagger \Phi_u)\Phi_d^\dagger \Phi_d + \lambda_7 (\Phi_d^\dagger \Phi_u)\Phi_u^\dagger \Phi_u ~+~ \text{h.c.} \Big]~, \nonumber
\end{eqnarray}
with $\Phi_u$ and $\Phi_d$ Higgs doublets with hypercharge $+1/2$.

The Higgs fields can be parametrized as
\begin{eqnarray}\label{eq:Higgs_fields}
\Phi_d &=& \begin{pmatrix} H_d^+ \\ \frac{1}{\sqrt{2}} (v_d + h_d + i a_d) \end{pmatrix} ~, \nonumber \\
\Phi_u &=& \begin{pmatrix} H_u^+ \\ \frac{1}{\sqrt{2}} (v_u + h_u + i a_u) \end{pmatrix} ~.
\end{eqnarray}

Assuming CP conservation, the mass eigenstates are given by
\begin{eqnarray}
\begin{pmatrix} h \\ H \end{pmatrix} &=& \begin{pmatrix} \cos{\alpha} & -\sin{\alpha} \\ \sin{\alpha} & \cos{\alpha} \end{pmatrix} \begin{pmatrix} h_u \\ h_d \end{pmatrix} ~, \nonumber \\
\begin{pmatrix} G \\ A \end{pmatrix} &=& \begin{pmatrix} \sin{\beta} & \cos{\beta} \\ \cos{\beta} & -\sin{\beta} \end{pmatrix} \begin{pmatrix} a_u \\ a_d \end{pmatrix} ~, \nonumber \\
\begin{pmatrix} G^\pm \\ H^\pm \end{pmatrix} &=& \begin{pmatrix} \sin{\beta} & \cos{\beta} \\ \cos{\beta} & -\sin{\beta} \end{pmatrix} \begin{pmatrix} H_u^\pm \\ H_d^\pm \end{pmatrix} ~,
\end{eqnarray}
where we define the basis-dependent ratio $\tan{\beta}~\equiv~v_u/v_d$~\footnote{In the following, we will restrict our attention to Type-II 2HDMs, for which $\tan\beta$ is a well defined basis-independent quantity~\cite{Haber:2006ue}.} and $\sqrt{v_d^2 + v_u^2} = 246$ GeV. $G^\pm$ and $G^0$ are Goldstone bosons, $A$ the pseudoscalar, $H^\pm$ the charged Higgs, and $h,H$ the light and heavy CP-even scalar, respectively. In the following, we will choose to work in the alignment limit, $\sin({\beta-\alpha})=1$, where the $h$ couplings are SM-like.

In the MSSM, the masses of the heavy scalars are clustered around a similar scale, with splittings arising only from small $D$-terms. However, in a general 2HDM framework, we have more freedom to get more sizable splittings. As a result, we can split the pseudoscalar mass such that $m_A < m_H \sim m_{H^+}$, as needed for a model that can fit the GCE (see Sec.~\ref{GCE}). In particular, we can write the mass eigenvalues as functions of the quartic couplings in Eq.~(\ref{eq:Higgs_potential}):
\begin{eqnarray}\label{eq:mCminusmA}
m_{H^\pm}^2-m_A^2=\tfrac{v^2}{2}(\lambda_5-\lambda_4)
\end{eqnarray}
and for $m_A\gg \lambda_i v$ and $\tan\beta\sim1$, 
\begin{eqnarray}
&&m_{H}^2-m_A^2\simeq\tfrac{v^2}{4}(\lambda_1+\lambda_2-2(\lambda_3+\lambda_4-\lambda_5)), \\ 
&&m_h^2\simeq\tfrac{v^2}{4}(\lambda_1+\lambda_2+2(\lambda_3+\lambda_4+\lambda_5+2\lambda_{67})),
\end{eqnarray}
while for  $m_A\gg \lambda_i v$ and $\tan\beta\gg 1$,
\begin{eqnarray}
m_{H}^2-m_A^2\simeq   \lambda_5 v^2, ~ ~ ~  m_h^2 \simeq  \lambda_2 v^2.   \label{eq:mh}
\end{eqnarray}
where $\lambda_{67}\equiv \lambda_6+\lambda_7$. From these expressions, it is clear that a splitting as large as $\sim$100 GeV between the pseudoscalar and the charged Higgs and between the pseudoscalar and the heavy Higgs $H$ can be obtained for $\lambda_4,\lambda_5=\mathcal O(1)$.

Electroweak precision measurements, and in particular the $\rho$ parameter, can give important constraints on spectra with large splittings. A notable exception are alignment models with $m_H\sim m_{H^\pm}$, but with an arbitrary large mass splitting between the pseudoscalar and the charged Higgs, which leads to a very small correction to the $\rho$ parameter. Further constraints arise from the requirement of vacuum stability and perturbativity of the quartic couplings. In Appendix~\ref{app:Higgsspectrum}, we will give examples of viable quartic couplings that are able to produce the spectra that we use in Sec.~\ref{GCE}, including the right mass for the SM-like Higgs boson discovered at the LHC.

Finally, additional constraints on the scalar spectrum come from flavor transitions. In general, both Higgs doublets can couple to up and down quarks, as well as to leptons, with generic Yukawa couplings. However, the most general Yukawa couplings lead to excessively large contributions to flavor changing neutral transitions. This has led to the various well-known ``Types'' of 2HDMs, in which different discrete $\mathbb{Z}_2$ symmetries are imposed; the two Higgs doublets have different $\mathbb{Z}_2$ charge, thus forbidding some of the Yukawa couplings. This condition also reduces the number of free parameters in Eq.~(\ref{eq:Higgs_potential}), since the $\mathbb{Z}_2$ symmetry demands $\lambda_6=\lambda_7=0$\footnote{We work under the assumption that the $\mathbb{Z}_2$ symmetry is softly broken by the mass term $B\mu(\Phi_d^\dagger \Phi_u)+\text{h.c.}$.}. In the following, we will focus on a Type-II 2HDM, for which the $\mathbb{Z}_2$ symmetry allows the doublet $\Phi_u$ to only couple to right-handed up quarks and the doublet $\Phi_d$ to only couple to right-handed down quarks and leptons. This type of 2HDM is interesting, since the MSSM at tree level is a particular limit of Type-II 2HDMs.


\section{Collider signals \label{sec:collider} }

In this section we present model-independent collider bounds on the new Higgs sector coupled to DM. We will keep the discussion as general as possible, considering a single new pseudoscalar and dark matter particle with coupling as in Eq.~(\ref{eq:Lsimple}).  Collider signals of simplified models of dark matter coupled to new scalars have also been discussed recently in~\cite{Harris:2014hga,Buckley:2014fba}.  For the set of models we consider in this paper, there will also be a rich set of collider signals associated with new charged states, such as the charged Higgs and the additional charged fermion states. We will discuss these in Section~\ref{themodel}.

\subsection{Present bounds \label{sec:Presentbounds}}

\begin{figure}[t]
\begin{center}
\includegraphics[width=0.49\textwidth]{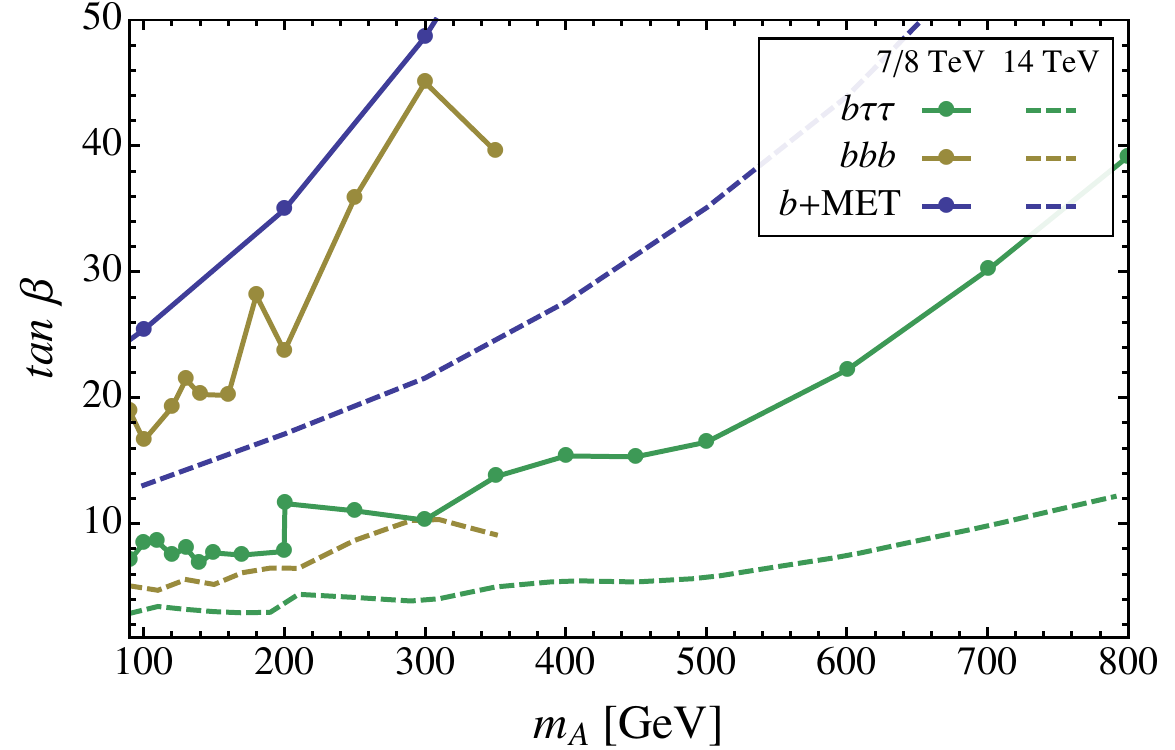}
\caption{Comparison of present LHC bounds for a heavy pseudoscalar in the visible and invisible decay channels, in each case assuming the maximum possible branching ratio arising in a Type-II 2HDM (solid lines). Upper limits on $\tan \beta$ are shown as a function of the new scalar mass $m_A$. 
Note that for $A \to b \bar{b}$ only 7 TeV limits are available, while all the others searches have been performed with the full 8 TeV, 20/fb data set. We also show extrapolated limits for a 14 TeV LHC with 300/fb (dashed lines), as explained in Section~\ref{sec:14tev}. 
\label{fig:Asearches}
}
\end{center}
\end{figure}

In a Type-II 2HDM, the $A$ coupling to down-type quarks, and charged leptons, is given by $y_f \tan \beta$, where $y_f=m_f/v$, with $m_f$ the mass of the corresponding fermion.  Assuming a Majorana fermion DM $\chi$ with interaction as in Eq.~(\ref{eq:Lsimple}), the invisible decay width of the heavy pseudoscalar is given by
\begin{equation}
	\Gamma(A \to \chi \chi)  = y_\chi^2 \frac{m_A}{4\pi} \sqrt{1 - \frac{4 m_\chi^2}{m_A^2} }~,
\end{equation}
while the visible decay width is
\begin{equation}
	\Gamma(A \to f \bar{f})  = n_c y_f^2\tan^2\beta \frac{m_A}{ 8 \pi} \sqrt{1 - \frac{4 m_f^2}{m_A^2} }~,
\end{equation}
where $n_c$ is the number of colors of the final state SM fermion. Depending on the values of $y_\chi$ and $\tan\beta$, different LHC search channels can play a role.

In Fig.~\ref{fig:Asearches} we compare the constraints on the visible and invisible $A$ decay from current LHC searches.  In the same figure, we also give 14 TeV projections for each of these searches, as discussed in Section~\ref{sec:14tev}. In each case, we assume the maximal possible branching ratio in order to show the optimal sensitivity that could be achieved in each channel. More precisely, in the case of the pseudoscalar decaying to bottom quarks and $\tau$ leptons, we assume $y_\chi=0$ and in the case of the pseudoscalar decaying invisibly, we assume BR$(A\to\chi\chi) \rightarrow 1$. For simplicity, results are shown in the narrow width limit for the pseudoscalar.

In the absence of invisible decays, in a Type-II 2HDM at large $\tan \beta$, the branching ratio of $A$  to $\tau \tau$ is about 10$\%$ and the one to $b \bar{b}$ about 90$\%$. The most stringent LHC constraints on $A$ come from heavy MSSM Higgs boson searches. The main production modes of $A$ are in association with $b$-quarks and in gluon fusion with a heavy quark loop. For $A \to b \bar{b}$,  we use the 7 TeV analysis from CMS \cite{Chatrchyan:2013qga}. For $A \to \tau \tau $, we take the 8 TeV results from ATLAS \cite{Aad:2014vgg}, which gives slightly stronger limits at high $m_A$ compared to the CMS analysis \cite{Khachatryan:2014wca}.  In order to derive conservative constraints, we assume that the heavy CP-even Higgs is somewhat heavier than the pseudoscalar, so that only the pseudoscalar contributes to the invisible or visible signature. This is different than the typical models used by the LHC experimental collaborations to interpret heavy Higgs  searches~\cite{Chatrchyan:2013qga,Aad:2014vgg,Khachatryan:2014wca}, in which $m_A=m_H$ and both the pseudoscalar and heavy CP-even Higgs contribute to the signature.

The $A \to \chi \chi$ decay mode gives rise to a $b+ \missET$ (mono-$b$) signal in $b$-associated production. To derive constraints, we simulate the signal with  {\sc{MadGraph5}} \cite{Alwall:2011uj},  {\sc{PYTHIA}} \cite{Sjostrand:2006za} and {\sc{DELPHES}} \cite{deFavereau:2013fsa}, and compare with the signal region SR1 in the ATLAS analysis \cite{Aad:2014vea}.  As Fig.~\ref{fig:Asearches} shows, the $\tau \tau$ channel gives the strongest constraint. This is the only channel for which a reconstruction of $m_A$ with $m_{\tau \tau}$ is possible, and furthermore can have smaller backgrounds if a leptonic $\tau$ decay is considered.

\begin{figure}[tb]
\begin{center} 
\includegraphics[width=0.48\textwidth]{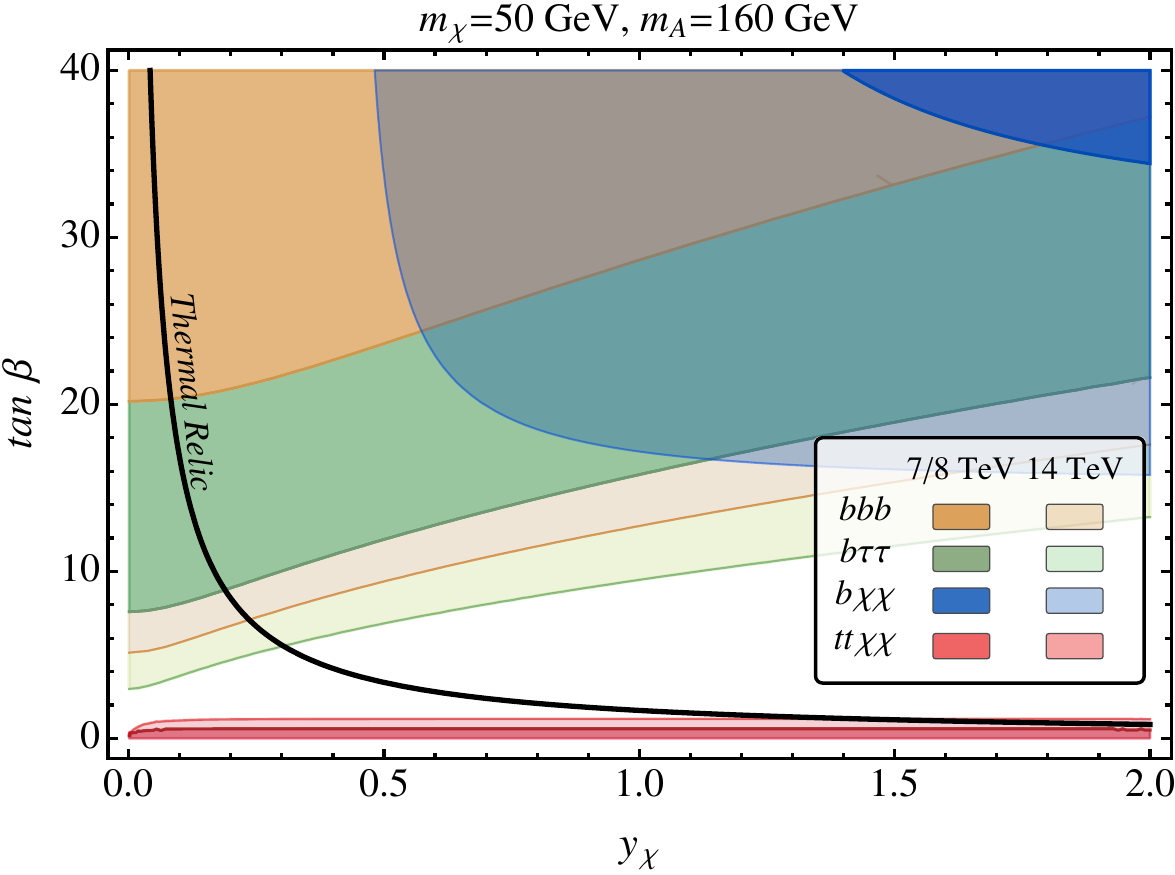}
\caption{Present and projected collider reach on the pseudoscalar couplings for a specific choice of DM and pseudoscalar masses. The parameters chosen are those favored for explaining the GCE, as discussed in Section~\ref{GCE}. We also show the 8 TeV limits from the $t \bar t + \missET$ search \cite{Zhou:2014dba}.
\label{fig:Aparamspace}
}
\end{center}
\end{figure}

Fig.~\ref{fig:Asearches} assumed maximal branching ratios in each case. The relative importance of the constraints is modified when both invisible and visible decays have sizable branching ratios. Next, we consider the interplay of these searches, as well as the requirement of a thermal relic. For concreteness, we focus on DM with mass $m_\chi = 50$ GeV, where the dominant annihilation channel is through a pseudoscalar of mass $m_A = 160$ GeV, such as that favored by the GCE (see Section~\ref{GCE}).

The constraints on the two free parameters $y_\chi$ and $\tan \beta$ are shown in Fig.~\ref{fig:Aparamspace}.  For both visible and invisible decays in $b$-associated production, a minimum $\tan \beta$ can be reached since the production mechanism relies on the $A b \bar{b}$ coupling, which is $\tan\beta$ enhanced. Furthermore, as expected, the mono-$b$ search covers the large $y_\chi$ region, which is difficult to probe with the $bbb$ search. In Fig.~\ref{fig:Aparamspace}, we also show limits at low $\tan\beta$ coming from the search for a pseudoscalar decaying invisibly and produced in association with top quarks \cite{Zhou:2014dba}. This search is able to exclude values of $\tan\beta \sim 0.5$, above which the production cross section becomes too small\footnote{In principle, invisible $h$ decay should also be included which would strengthen constraints, but the relation of its invisible decay to the $A$ invisible decay is more model-dependent, so we do not include it in this figure.}. From the figure, we can see that the $t\bar t + \missET$ and $b \tau\tau$ signatures are complementary to probe the region of parameter space where DM is a thermal relic. The $b+\missET$ search, instead, is not effective in this region.

Finally, in this context, it is interesting to understand where the SM-like Higgs boson discovered by the LHC is placed. The most stringent direct bound on the branching ratio of the Higgs decaying invisibly comes from the CMS analysis~\cite{Chatrchyan:2014tja} that combines the searches for a Higgs produced in vector-boson-fusion and for a Higgs produced in association with a $Z$ boson. This analysis is able to put a bound on the branching ratio into invisible at  $58\%$ at the $2\sigma$ level.  Additional bounds come from searches for a Higgs produced in association with tops and decaying invisibly. Following the results in Ref.~\cite{Zhou:2014dba}, the branching ratio into invisible is bounded at $40 \%$. Using this latter bound, the SM Higgs coupling to DM is then constrained to be about 0.02, with which a thermal relic is possible in only-Higgs mediated models when $m_\chi$ is very close to $m_h/2$~\cite{deSimone:2014pda}. This is a much stronger bound than we get from the $b + \missET$ signature, which is only able to constrain at best a coupling of $y_\chi \sim 1$ for the same scalar mass.

\subsection{LHC Run 2 projections \label{sec:14tev}}

In Figs.~\ref{fig:Asearches}-\ref{fig:Aparamspace}, we also show projections for the 14 TeV LHC with 300/fb integrated luminosity. 

For the mono-$b$ limits, we simulate events with the cuts in \cite{Lin:2013sca}, taking advantage of the background simulation already done there. A full projection for the visible channels requires modeling of the gluon fusion to $A$ processes, as well as the SM $bbb$,  $\tau \tau$ and $b \tau \tau$ backgrounds. A detailed collider study is beyond the scope of the current paper. We present a simple estimate for these channels. We extrapolate both signal and background based on pdf luminosity, assuming for the signal on-shell production of a new heavy mediator. Our method is similar to that in~\cite{ColliderReach}, but we have also to include the $b-$quark PDF in our analysis. Combining these results gives an estimate for the improvement in $S/\sqrt{B}$, which we use to determine the improvement in the $\tan \beta$ bound for each given $m_A$. For this reason, in Fig.~\ref{fig:Asearches} the projection for the $bbb$ search is only shown up to $m_A = 350$ GeV. In all cases, for $A$ production, we consider only $b$-associated production. The gluon fusion channel will become important at low values of $\tan \beta$. For the values of $\tan\beta$ probed at 8 TeV, we have checked that gluon fusion is subdominant if compared to $b$-associated production, with the explicit cross section limits on $gg \to bbA$ given in Refs.~\cite{Aad:2014vgg,Khachatryan:2014wca}. Therefore, the projections for 14 TeV can be regarded as conservative, in that they ignore the additional constraining power from gluon fusion.

Two comments on the figures are in order: projections of the $bbb$ and $b \tau\tau$ channels show that, as expected, the bound on $\tan\beta$ from the $A\to bb$ decay will be weaker than the bound from $A\to\tau\tau$. Still, the difference in the reach of the two channels is not that large, especially at low values of $m_A$, motivating the search for $A\to bb$ with additional data beyond the 5/fb 7 TeV set. Additionally, as we can see from Fig.~\ref{fig:Aparamspace}, improved constraints on the $b \tau\tau$ ($bbb$) and $tt+$ MET channels from LHC run 2 could help cover much of the region of parameter space where DM is a thermal relic.

\section{Extended Higgs portal models \label{sec:scalarDM} }

Models of scalar DM coupled to an extended Higgs sector have been studied extensively in the literature, including the case with a 2HDM Higgs content \cite{Drozd:2014yla,He:2011gc,He:2013suk,Bai:2012nv,Wang:2014elb}. Especially for light DM, this is a much more constrained scenario than for fermion DM, as we review in the following.

If DM is assumed to be a real scalar, $S$, and a gauge singlet under the SM, the lowest dimension gauge invariant operator that allows direct couplings to the SM Higgs is \cite{Burgess:2000yq}
\begin{align}
\label{HiggsPortal}
\mathcal{L}_\text{DM} \supset \lambda_S \Phi^\dagger \Phi S^2
~,
\end{align}
where $\Phi$ is the SM Higgs doublet. However, for DM mass $m_S \lesssim m_h/2$ the large values of $\lambda_S$ needed to sufficiently deplete the abundance of $S$ before freeze-out are in strong conflict with constraints from the invisible width of the SM Higgs. For heavier masses, aside from a small window of viable parameter space very close to the Higgs resonance, LUX \cite{Akerib:2013tjd} rules out thermal DM for $m_S \lesssim 100$ GeV \cite{deSimone:2014pda} (see also~\cite{Feng:2014vea}, for a recent analysis). However, this does not have to be the case if the model is slightly extended beyond the simple Lagrangian of Eq.~(\ref{HiggsPortal}).

A concrete example is the singlet scalar extension of a 2HDM. Once again, the singlet scalar $S$ is the DM candidate, but it now possesses interactions with $\Phi_{d,u}$ through the terms
\begin{align}
\mathcal{L}_\text{DM} \supset S^2 \left[ \lambda_d \Phi_d^\dagger \Phi_d + \lambda_u \Phi_u^\dagger \Phi_u + \lambda_{du} \big( \Phi_d^\dagger \Phi_u + \text{h.c.} \big)  \right]
~.  \nonumber
\end{align}
The introduction of a second Higgs doublet allows $S$ to annihilate through an s-channel $H$ as well. For CP-conserving interactions, annihilation through $A$ is not allowed. As a result, it is possible to slightly uncouple the annihilation rate from the coupling of the SM Higgs with DM and thus ease the tension with the invisible Higgs and LUX constraints to some degree.
However, doing so can require tunings in the couplings above, in order to sufficiently suppress the interaction strength with the SM Higgs. Furthermore, since the heavy $H$ also leads to spin-independent (SI) nucleon scattering, thermal DM below 100 GeV is still very restricted by direct detection constraints. More detailed analysis of the model can be found in Ref.~\cite{Drozd:2014yla}.

Another option is to consider fermion DM. We begin by summarizing the constraints on a model with the Higgs sector given just by the SM Higgs. Light thermal DM is particularly difficult to achieve. In particular, a strictly weakly interacting fermion DM candidate is in strong tension with the null results of current direct detection experiments. For example, a DM candidate like a 4th generation heavy Dirac neutrino has $Z$ couplings that would lead to a nucleon elastic scattering rate many orders of magnitude beyond what is currently allowed by LUX (see e.g. \cite{deSimone:2014pda}).

A fix for this issue is to introduce a new gauge singlet fermion that is allowed to mix with the active components of the DM after electroweak symmetry breaking. This can be explicitly realized by coupling the sterile and active components through renormalizable interactions involving the SM Higgs.  Dubbed  ``singlet-doublet" DM \cite{Enberg:2007rp,Cohen:2011ec,Cheung:2013dua}, the simplest possibility is to introduce a Majorana gauge singlet and a vector-like pair of $SU(2)$ doublets.
Since the mixing of the DM candidate originates from Yukawa interactions with the SM Higgs, such scenarios generically are quite constrained by spin-independent  direct detection limits, and for light DM this usually necessitates living somewhat near ``blind-spots" of the theory, where the coupling to the SM Higgs is strongly suppressed. Furthermore, since annihilation through the SM Higgs is velocity suppressed, to accommodate a thermal relic, DM annihilations involving gauge bosons need to be the dominant channel, and for DM lighter than $\sim 80$ GeV, this demands a strong coupling to the $Z$. However, precision electroweak measurements near the $Z$-pole strongly constrain such a coupling when $m_\chi \lesssim m_Z/2$ \cite{ALEPH:2005ab}. Hence, additional annihilation mediators would greatly benefit the prospects for light thermal DM.

In light of this reasoning, in the next section we will study a model that adds a second Higgs doublet to this simple fermion DM scenario. The singlet-doublet case discussed above is one particular limit of this model. As we will see, the extension to a 2HDM vastly alleviates the tension of $\lesssim \mathcal{O}(100)$ GeV singlet-doublet DM with current experimental constraints.  


\begin{table}[t]
\centering
\begin{tabular}{| c || c | c | c |}
\hline
Field & Charges & Spin \\ \hline \hline
$\Phi_d$ & $(\mathbf{1}, \mathbf{2}, +1/2)$ & 0 \\ \hline
$\Phi_u$ & $(\mathbf{1}, \mathbf{2}, +1/2)$ & 0 \\ \hline
$S$ & $(\mathbf{1}, \mathbf{1}, 0)$ & 1/2 \\ \hline
$D_1$ & $(\mathbf{1}, \mathbf{2}, -1/2)$ & 1/2 \\ \hline
$D_2$ & $(\mathbf{1}, \mathbf{2}, +1/2)$ & 1/2 \\ \hline
\end{tabular}
\caption{Field content of our model with fermion DM. The given charges are for the SM gauge groups $SU(3)\times SU(2)\times U(1)_Y$.}
\label{fields}
\end{table}

\section{Fermion DM in a 2HDM}
\label{themodel}

We define a simple UV-complete extension of the SM, whose scalar sector is that of a  Type-II 2HDM, and whose DM sector consists of one gauge singlet Weyl fermion, $S$, and two Weyl $SU(2)$ doublets $D_1$ and $D_2$, oppositely charged under $U(1)_Y$, as shown in Table \ref{fields}. We assume that the fermionic fields of the dark sector are $\mathbb{Z}_2$ odd, ensuring the stability of the lightest state. 

The DM sector Lagrangian has the terms
\begin{align}
\mathcal{L}_{\text{DM}} \supset &-\frac{1}{2} M_S S^2 - M_D D_{1} D_{2} 
\nonumber \\
& - y_1 S D_{1} \Phi_{1} - y_2 S \Phi^\dagger_2 D_2 + \text{h.c.}
~,
\label{DarkLagrangian}
\end{align}
where 2-component Weyl and $SU(2)$ indices are implied. The notation $\Phi_{1,2}$ indicates that we will allow different permutations for which Higgs doublets (down or up-type) couple to $D_1$ and $D_2$. From here forward, we will use the naming scheme where $\Phi_1 = \Phi_d$, $\Phi_2 = \Phi_u$ will be called the ``\emph{du}" model, and similarly for other choices of $\Phi_1$ and $\Phi_2$. The naming schemes are given in Table~\ref{modelnames}. Note that in Eq.~(\ref{DarkLagrangian}), we have made the simplifying assumption that $D_1$ and $D_2$ each only couple to a single Higgs doublet. In the case of the $dd$ and $uu$ models, this can be enforced by the $\mathbb{Z}_2$ symmetry of the 2HDM scalar sector (see Sec.~\ref{2HDM}), that is assumed to be broken only by mass terms. This is not the case for the $du$ and $ud$ models. Still, our qualitative conclusions would not change significantly if all possible Yukawas are allowed to be of comparable strength.

\begin{table}[t]
\centering
\begin{tabular}{| c || c |}
\hline
Name & Model \\ \hline \hline
\emph{du} & $\Phi_1 = \Phi_d$, $\Phi_2 = \Phi_u$ \\ \hline
\emph{ud} & $\Phi_1 = \Phi_u$, $\Phi_2 = \Phi_d$ \\ \hline
\emph{dd} & $\Phi_1 = \Phi_d$, $\Phi_2 = \Phi_d$ \\ \hline
\emph{uu} & $\Phi_1 = \Phi_u$, $\Phi_2 = \Phi_u$ \\ \hline
\end{tabular}
\caption{Naming scheme for Eq.~(\ref{DarkLagrangian}).}
\label{modelnames}
\end{table}

The situation where only the SM Higgs doublet appears in the interactions of the Lagrangian (\ref{DarkLagrangian})  is the singlet-doublet model discussed in Section~\ref{sec:scalarDM} and  can be identified as the limiting case of $\tan{\beta} \gg 1$ in the \emph{uu} model. In this paper, we do not perform an analysis of this model, as it has already been covered in great detail in \cite{Cohen:2011ec} and \cite{Cheung:2013dua}. Furthermore, since the \emph{du} and \emph{ud} model can be mapped into each other by a simple switch of $y_1 \leftrightarrow y_2$, only one systematic study of the two is needed.

Although the $du$ model embodies the spirit of bino-Higgsino DM in the MSSM, it allows much more freedom since we are freed from restrictions such as the Yukawa couplings being fixed by the gauge sector, and compared to singlino-Higgsino DM in the NMSSM, the model requires fewer new degrees of freedom\footnote{Bino-Higgsino mixing of the MSSM can be related through the identifications $y \leftrightarrow g'$, $\tan{\theta} \leftrightarrow -1$, $M_S \leftrightarrow M_1$, and $M_D \leftrightarrow \mu$, where $g'$ is the hypercharge gauge coupling, $M_1$ is the bino soft SUSY breaking mass, and $\mu$ is the Higgsino mass term. For singlino-Higgsino mixing of the NMSSM, the appropriate identifications are $y \leftrightarrow \sqrt{2} \lambda$, $\tan{\theta} \leftrightarrow +1$, $M_S \leftrightarrow \mu' + 2 \kappa s$, and $M_D \leftrightarrow \lambda s$, where $\lambda$ is the Yukawa coupling for the trilinear singlet-Higgs interaction of the superpotential, $\mu'$ is the supersymmetric mass term for the singlet, $\kappa$ is the Yukawa trilinear self-interaction for the singlet, and $s$ is the VEV of the singlet scalar.}. For simplicity, we take all couplings to be real, and work in the convention $M_S, M_D > 0$. Keeping the sign of $M_D$ and $M_S$ fixed, we see that we can parity transform just $S \to -S$ or both $D_1 \to -D_1$ and $D_2 \to -D_2$.  Either of these choices in field re-definitions results in simultaneously flipping the sign of $y_1$ and $y_2$. However, their relative sign remains unchanged. Therefore, the only physical sign in our couplings is that of $y_2/y_1$. Hence, we will often trade the couplings $y_1$, $y_2$ in favor of $y$ and $\tan{\theta}$ defined by 
\begin{equation}
y_1 \equiv y \cos{\theta} ~,~ y_2 \equiv y \sin{\theta}
~.
\end{equation}

The DM doublets are parametrized as
\begin{equation}
D_1 = \begin{pmatrix} \nu_1 \\ E_1 \end{pmatrix} \quad , \quad D_2 = \begin{pmatrix} -E_2 \\ \nu_2 \end{pmatrix} ~ ,
\end{equation}
where $\nu_{1,2}$, $E_{1,2}$ are the neutral, charged components  of the doublets, respectively. 
$E_1$ and $E_2$ combine to form an electrically charged Dirac fermion of mass $M_D$. 
To avoid constraints coming from chargino searches at LEP, we only consider values of $M_D \gtrsim 100$ GeV \cite{LEP:Chargino}.

After electroweak symmetry breaking, one ends up with three Majorana mass eigenstates, in general all mixtures of the singlet $S$ and the neutral components ($\nu_1$ and $\nu_2$) of the $SU(2)$ doublets. In the $(S, \nu_1, \nu_2)$ basis, the neutral mass matrix for the fermionic dark sector is
\begin{equation}
M_\text{neutral} = \begin{pmatrix} M_S & \frac{1}{\sqrt{2}} y_1 v_1 & \frac{1}{\sqrt{2}} y_2 v_2 \\  \frac{1}{\sqrt{2}} y_1 v_1& 0 & M_D \\ \frac{1}{\sqrt{2}} y_2 v_2 & M_D & 0 \end{pmatrix} 
~,
\label{MassMatrix}
\end{equation}
where $v_{1,2}$ are the VEVs of $\Phi_{1,2}$. The lightest mass eigenstate of $M_\text{neutral}$ will be the stable Majorana DM candidate, which we write as a 2-component Weyl fermion $\chi_\alpha$. We will denote the $\chi_\alpha$ composition in terms of the gauge eigenstates as
\begin{equation}
\chi_\alpha = N_S^\chi S_\alpha + N_1^\chi \nu_{1 \alpha} + N_2^\chi \nu_{2 \alpha}
~.
\end{equation}

In most of the viable parameter space for thermal DM, especially for light DM $m_\chi \lesssim \mathcal{O}(100)$ GeV, it is usually the case that the $\chi$ is singlet-like and $ m_\chi \sim M_S \ll M_D$. This is because a large doublet component is strongly constrained by direct detection.  Therefore, it is useful to write down the approximate form of $N_{S,1,2}^\chi$ when both $M_S$ and $y_{1,2}v_{1,2}$ are much smaller than  $M_D$. In this limit, one finds
\begin{align} 
N_S^\chi \approx 1  ~,~~ N_1^\chi \approx & -\frac{1}{\sqrt{2}} \frac{y_2 v_2}{M_D} ~,~~ N_2^\chi \approx -\frac{1}{\sqrt{2}} \frac{y_1 v_1}{M_D}
~.  \label{MixingAngles} 
\end{align}

\subsection{Interactions}
\label{interactions}

We present the relevant DM interactions in this section. For the rest of the analysis, we will give our analytic expressions in terms of 4-component notation, since they can be written in a more compact form. We write the charged state as $E$, a Dirac fermion of electric charge $Q=-1$ and mass $M_D$.

The interactions of $\chi$ with $h$, $H$, $A$, and $H^+$  are
\begin{align}
\mathcal{L} &\supset \lambda_{\chi h} h \bar{\chi} \chi + \lambda_{\chi H} H \bar{\chi} \chi + \lambda_{\chi A} A \bar{\chi} i \gamma^5 \chi 
\nonumber \\
&+ \big[ H^+ \bar{\chi} \left( \lambda_{+ s} + \lambda_{+ p} \gamma^5 \right) E + \text{h.c.} \big].
\label{HiggsL}
\end{align}
The full forms for these couplings are given in Appendix~\ref{app:higgscouplings}. Note that in their explicit analytic form in Eqs.~(\ref{duHiggs})-(\ref{uuHiggs}), the Higgs couplings are proportional to $y^2$, reiterating that the only physical sign is that of $\tan{\theta}$. All of these couplings rely on the singlet-doublet mixing. Also from Eqs.~(\ref{duHiggs})-(\ref{uuHiggs}), it is important to note that, as long as $D_1$ and $D_2$ couple to the same Higgs doublet (i.e. the \emph{dd} and \emph{uu} models), there is a generic ``blind-spot" of the theory that is independent of $\tan{\beta}$: for  $(m_\chi~+~M_D \sin{2 \theta}) \sim 0$ the couplings to the CP-even scalars identically vanish. We will often be considering the case where $m_\chi \ll M_D$, since a large doublet component is strongly constrained by direct detection. In this limit, the blind-spot in these scenarios will necessitate $|\tan{\theta}| \ll 1$ or $|\tan{\theta}| \gg 1$, along with $\tan{\theta} < 0$. Furthermore, for  $m_\chi \ll M_D$, it is particularly simple to write the coupling of $\chi$ to the neutral Higgs bosons as
\begin{eqnarray}
\lambda_{\chi h} &\approx \frac{y_1 y_2}{2 M_D} \left(  N_1^h v_2 + N_2^h v_1 \right)
\nonumber \\
\lambda_{\chi H} &\approx \frac{y_1 y_2}{2 M_D} \left(  N_1^H v_2 + N_2^H v_1 \right)
\nonumber \\
\lambda_{\chi A} &\approx -\frac{y_1 y_2}{2 M_D} \left(  N_1^A v_2 - N_2^A v_1 \right)\,,
\end{eqnarray}
where $N_{1,2}^{h,H,A}$ are the projections of the physical Higgses into the gauge eigenstates, as defined in Eq.~(\ref{HiggsParam}), and are functions of $\tan\beta$.

Through mixing with $\nu_1$ and $\nu_2$, $\chi$ can also have interactions with the electroweak gauge bosons of the form
\begin{equation}
\label{eq:gaugeLagrangian}
\mathcal{L} \supset g_{\chi Z} Z_\mu \bar{\chi} \gamma^\mu \gamma^5 \chi + \big[ W_\mu^+ \bar{\chi} \gamma^\mu \left( g_{+ v} + g_{+ a} \gamma^5 \right) E + \text{h.c.} \big]
\end{equation}
such that
\begin{align}
g_{\chi Z} &= \frac{-g}{4 c_w} \left[ \left( N^\chi_1\right)^2 - \left( N^\chi_2\right)^2 \right] \approx \frac{g}{8 c_w M_D^2} \left( y_1^2 v_1^2 - y_2^2 v_2^2 \right)
\nonumber \\
g_{+ v} &= \frac{g}{2 \sqrt{2}} \left( N_1^\chi + N_2^\chi\right) \approx  \frac{-g}{4 M_D} \left( y_1 v_1 + y_2 v_2 \right)
\nonumber \\
g_{+ a} &= \frac{-g}{2 \sqrt{2}} \left( N_1^\chi - N_2^\chi\right) \approx  \frac{-g}{4 M_D} \left( y_1 v_1 - y_2 v_2 \right)~,
\label{GaugeCouplings}
\end{align}
where the latter approximate expressions hold for $m_\chi \sim M_S \ll M_D$. $g$ is the $SU(2)$ gauge coupling and $c_w$ is the cosine of the Weinberg angle. It is apparent that, for $M_S \ll M_D$, $g_{\chi Z}$ is small, and, furthermore, even for large $SU(2)$ doublet mixing, $g_{\chi Z}$ can vanish identically if $N_1^\chi = \pm N_2^\chi$, i.e. when $\tan{\theta} = \pm v_1/v_2$. For the sake of brevity, we have not written down the interactions of the heavier Majorana mass eigenstates of the theory. Although such terms are only relevant when the mass splittings are not too large, we do include them in our final analysis.

The analytic forms for the Higgs and gauge interactions we have written down so far are not completely general, since they depend on the sign of the mass terms for the physical states of the theory. In particular, if upon diagonalizing to the mass eigenstate basis we encounter states with a ``wrong sign" mass term, a field redefinition that preserves the Majorana character of the field must be performed, e.g. $\chi \to i \gamma^5 \chi$. This transformation preserves the canonical form of the kinetic terms, but may introduce additional $\gamma^5$'s or $i$'s in the interaction terms. These changes lead to physical consequences. The explicit analytic expressions written down in this paper must then be appropriately modified if this is indeed the case.

\begin{figure*}[t]
\begin{center}
\includegraphics[width=0.49\textwidth]{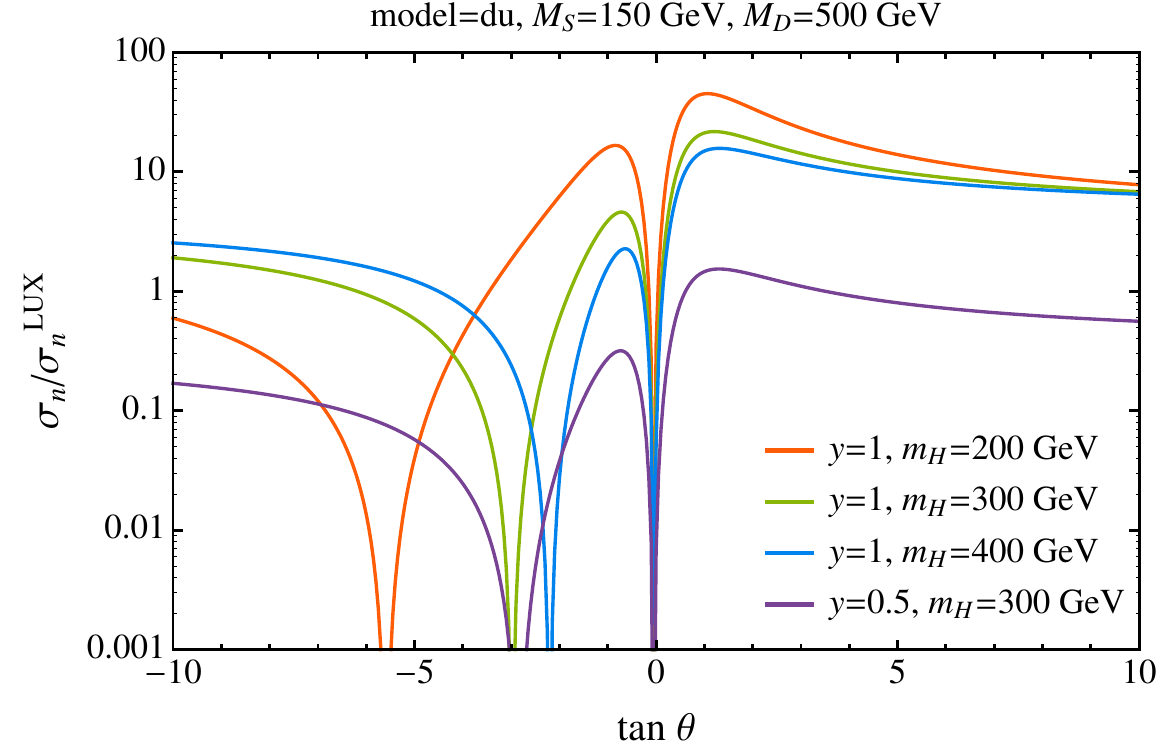}
\includegraphics[width=0.49\textwidth]{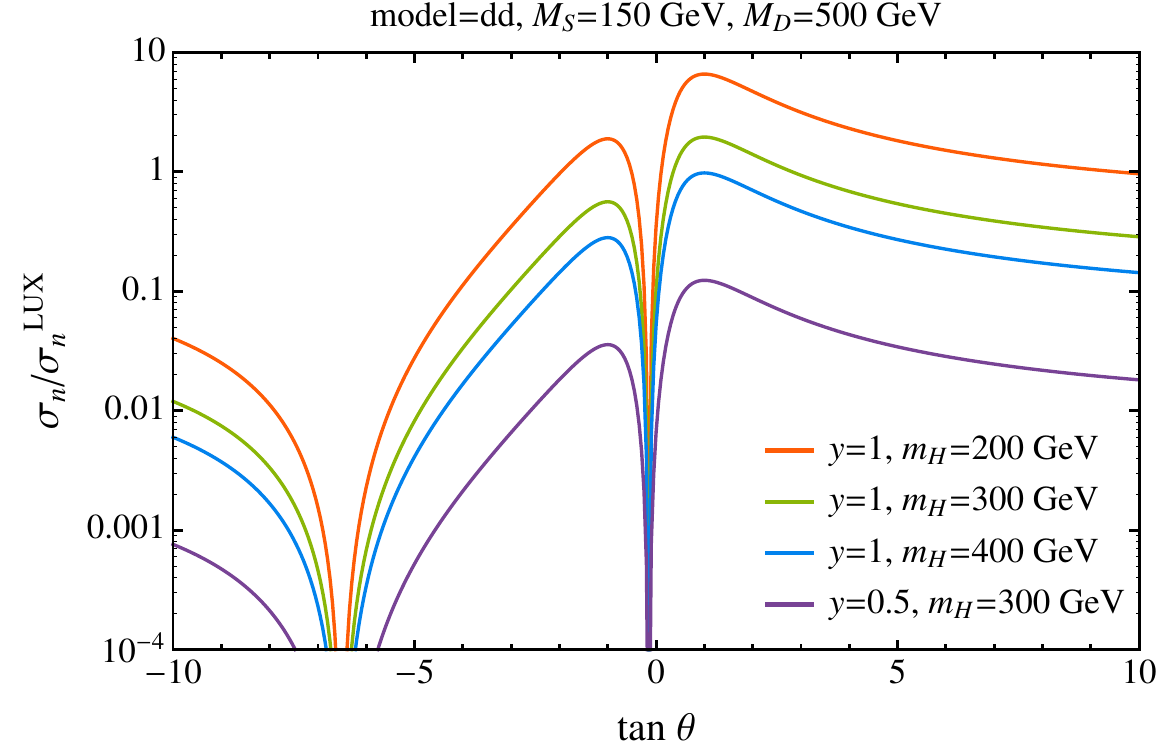}
\caption{Spin-independent tree-level direct detection cross section normalized to the LUX excluded cross section as a function of $\tan\theta$, for several values of $y$ and $m_H$, in the ({\emph{left}}) {\emph {du}} and ({\emph{right}}) {\emph {dd}} singlet-doublet models. The default parameters are $M_S = 150\ \GeV, M_D = 500\ \GeV,$ and $\tan \beta = 5$.}
\label{SIdudd}
\end{center}
\end{figure*}

\subsection{General Constraints}
\label{generalconstraints}

To be considered as a realistic DM candidate, $\chi$ must satisfy a handful of experimental constraints:
\begin{enumerate}
	\item The thermal relic density of $\chi$ must satisfy $\Omega h^2 \approx 0.1199 \pm 0.0027$, in agreement with current measurements from {\it WMAP} and {\it Planck}~\cite{Ade:2013zuv}. Furthermore, the annihilation rate must lie below the 95 \% CL  upper limits from gamma-ray observations of dwarf spheroidal satellite galaxies of the Milky Way~\cite{Ackermann:2015zua}. \label{enum:1}
	\item The scattering rate of $\chi$ with nuclei is below the current spin-independent and spin-dependent limits from LUX \cite{Akerib:2013tjd,Savage:2015xta}.\label{enum:2}
	\item If $m_\chi < m_h/2$, then BF$(h \to \chi \chi) \lesssim 19\%$, coming from global fits to the observed Higgs couplings \cite{Belanger:2013xza}. Note that our results do not change appreciably if we impose weaker bounds on the invisible Higgs decay at the level of $50-60\%$, as obtained from direct searches~\cite{Chatrchyan:2014tja,ATLAS:2013pma}. \label{enum:3}
	\item If $m_\chi < m_Z/2$, then $\Gamma (Z \to \chi \chi) \lesssim 2$ MeV, coming from LEP precision electroweak measurements near the $Z$-pole \cite{ALEPH:2005ab}. \label{enum:4}
	\item No other constraints from the LHC or other colliders are violated. In particular, we will consider LEP and LHC direct searches for a heavy Higgs and electroweakinos, and mono-$b$ constraints.
\end{enumerate}

\vspace{.5cm}
{\bf Relic Density.}
For DM masses $m_\chi \lesssim 80$ GeV, the dominant annihilation channels governing freeze-out and annihilation today are $s$-channel exchange of $Z$ and/or $A$ (if $A$ is sufficiently light). Since $\chi$ is Majorana, for $Z$-exchange the $s$-wave contribution is chirality suppressed by the mass of the final state SM fermions. Meanwhile, $A$-exchange is also suppressed by fermion mass but can be enhanced for large $\tan \beta$.

For more massive DM, specifically when $\chi$ is sufficiently heavier than $m_W$, $\chi$ can annihilate into pairs of $W^\pm$'s or $Z$'s, and both of these processes are in general unsuppressed by the $\chi$ relative velocity. Once $m_\chi$ is taken to be larger than a few hundred GeV, annihilations to pairs of scalars  become relevant, which depends on additional couplings in the full Higgs sector. We will therefore restrict our study to the parameter space where $\chi$ is sufficiently light such that final states including $H$, $A$, and $H^\pm$ do not contribute significantly to the calculation of $\Omega h^2$. With this simplifying assumption, we can safely ignore the heavy Higgs self-interactions present in the full 2HDM Lagrangian.

We find that in the calculation of $\Omega h^2$, resonances and co-annihilations can play an important role in setting the relic abundance in certain regions of parameter space.  Since a proper calculation of $\Omega h^2$ requires a careful treatment of such effects, we implement our model with {\tt FeynRules 2.0} \cite{Alloul:2013bka} and {\tt micrOMEGAS\_3.6.9} \cite{Belanger:2013oya} in the calculation of the relic abundance and numerically scan over the parameter space of our different models. In doing so, we have checked the output of the dominant annihilation channels analytically.  

\begin{figure*}[t]
\begin{center}
\includegraphics[width=0.49\textwidth]{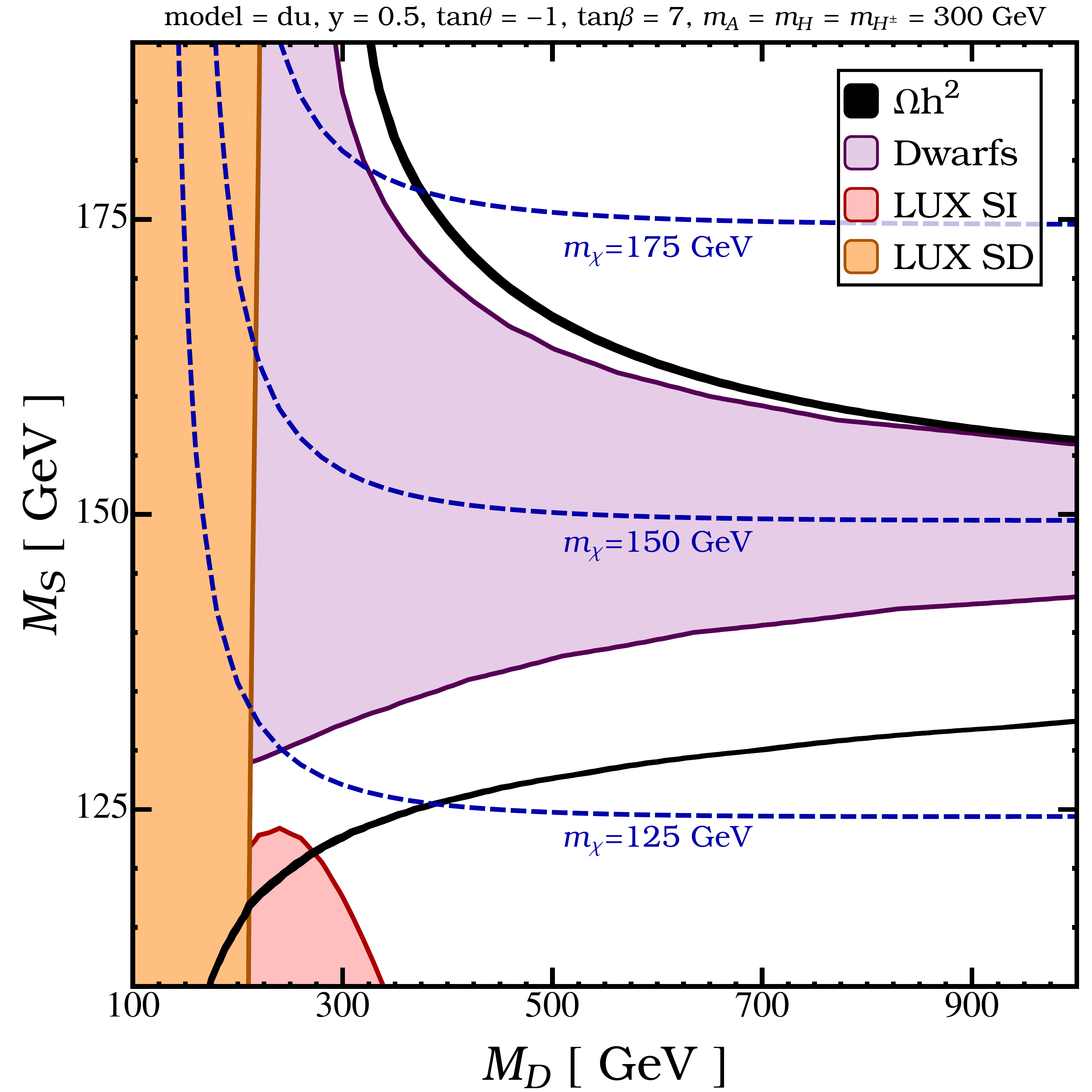}
\includegraphics[width=0.49\textwidth]{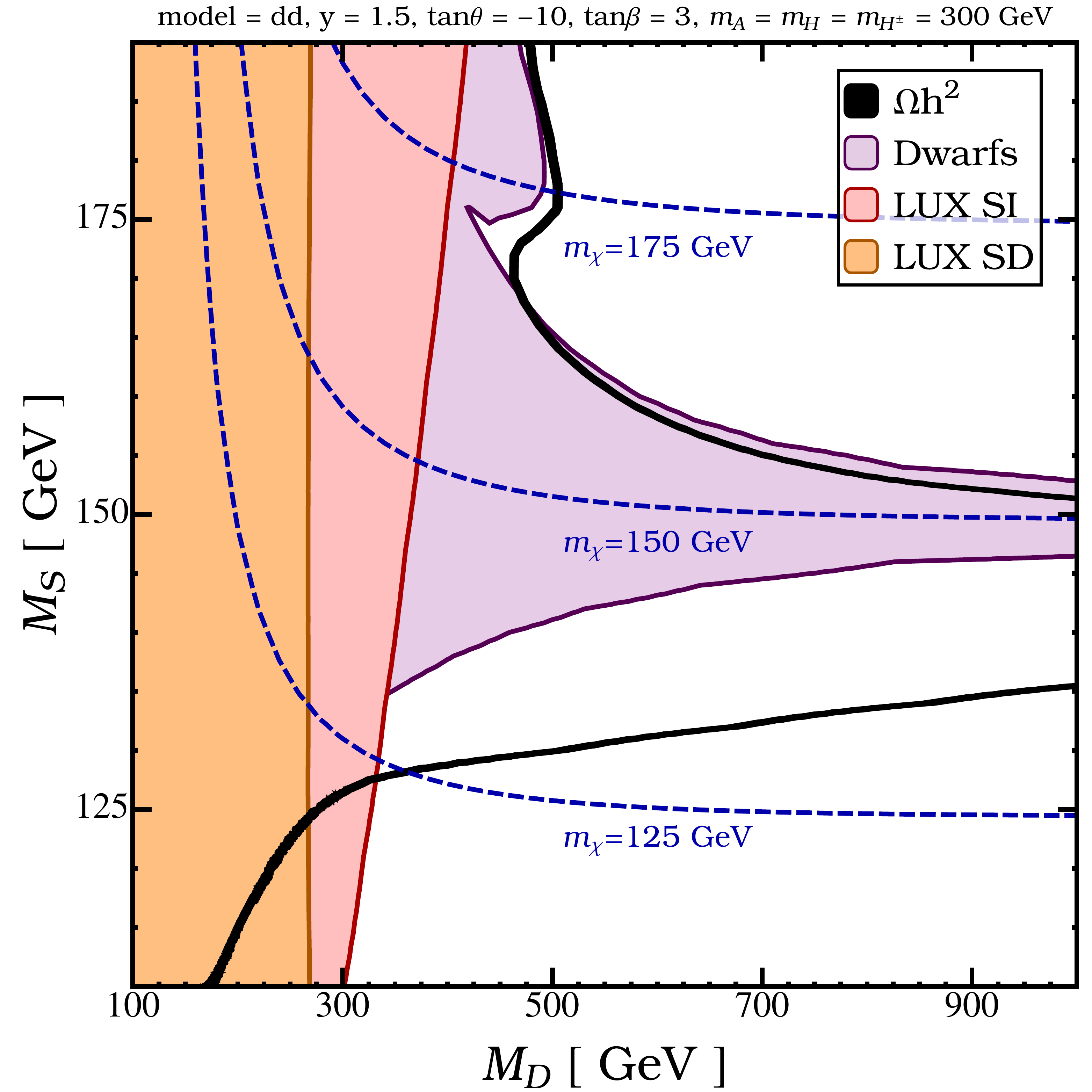}
\caption{Summary plot for the \emph{du} model for $y=0.5$, $\tan{\theta} = -1$, and $\tan{\beta}=7$ (\emph{left}) and for the \emph{dd} model for $y=1.5$, $\tan{\theta} = -10$, and $\tan{\beta}=3$ (\emph{right}). Shown are contours of $\Omega h^2 \sim 0.12$ (black) and $m_\chi$ (blue-dashed) and exclusion regions from direct detection searches at LUX for the spin-independent process (red) and spin-dependent process (orange). We also show regions that are above the upper limits from gamma-ray observations of dwarf galaxies (purple). In both cases, the masses of the heavy scalars are fixed to be $300$ GeV.
\label{fig:scan1} }
\end{center}
\end{figure*}

\begin{figure*}[tbh]
\begin{center}
\includegraphics[width=0.49\textwidth]{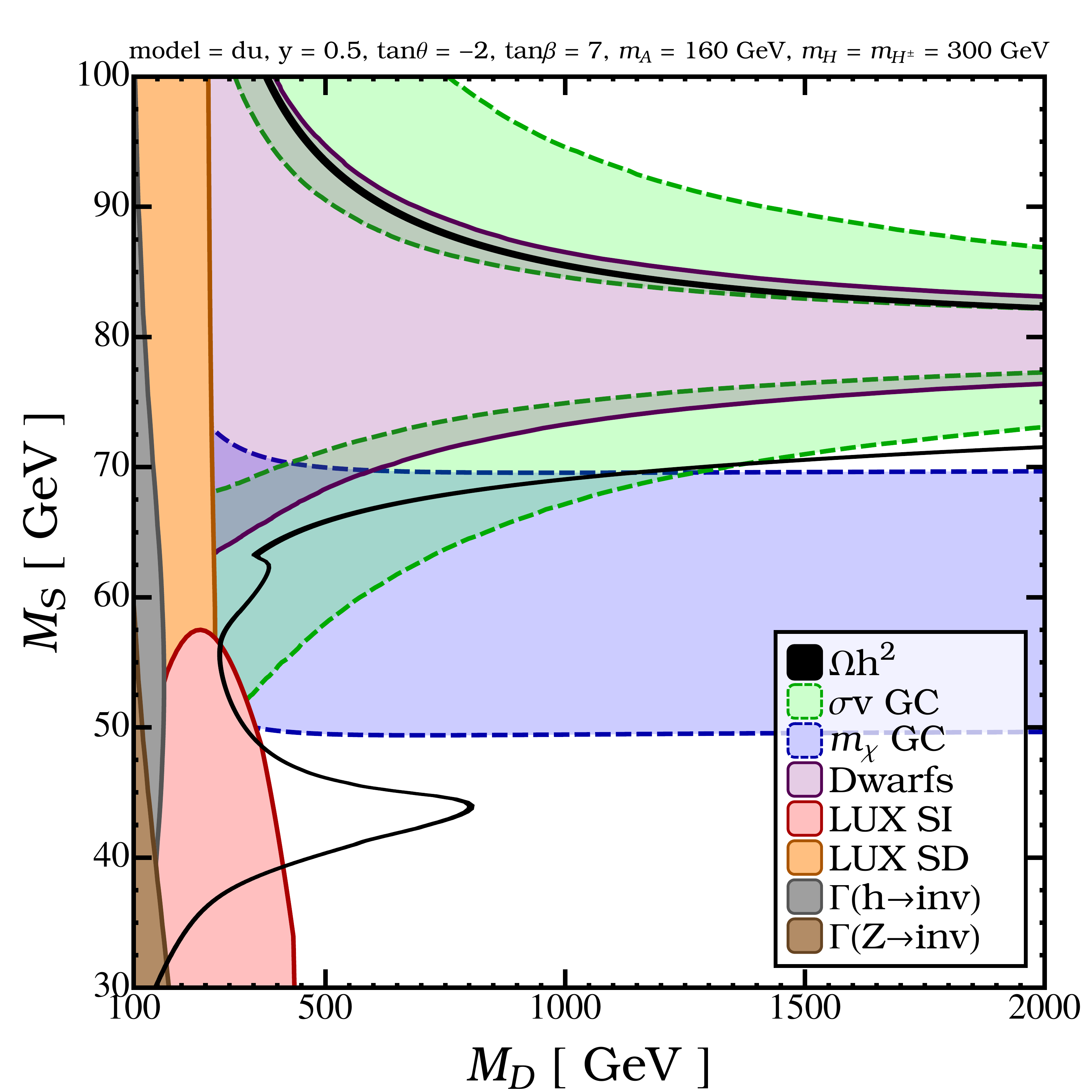}
\includegraphics[width=0.49\textwidth]{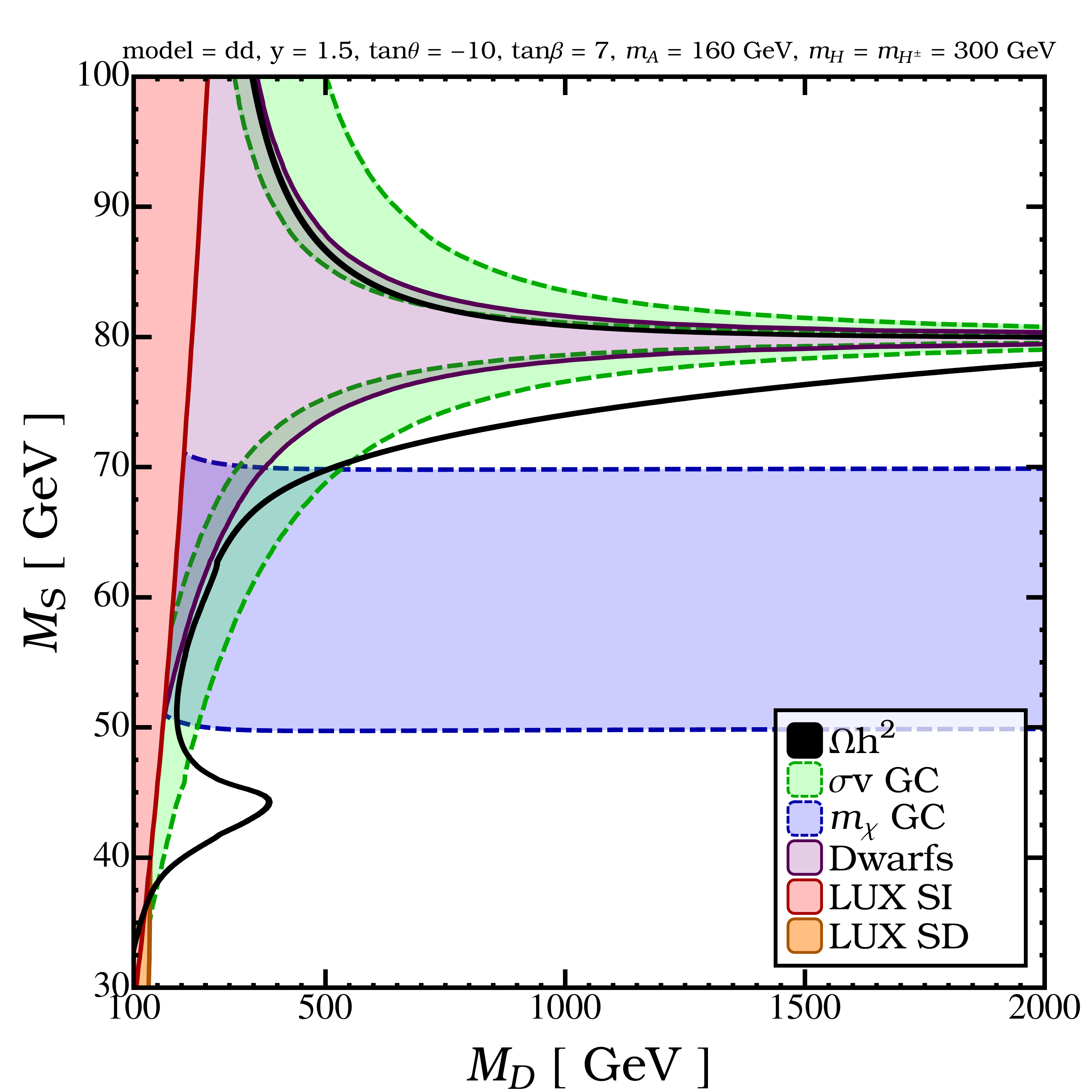}
\caption{Examples of viable parameter space for the GCE. In the left panel we show the \emph{du} model; in the right the \emph{dd} model. A DM with the correct relic density, mass, and $s$-wave annihilation cross section for the GCE is found when there is an overlap of the solid black curve, the blue dashed region, and the green dashed region. Regions are excluded by direct detection searches at LUX for the spin-independent process (red) and spin-dependent process (orange). In brown and gray we show the regions excluded by $Z\to$ invisible and by $h\to $invisible, respectively (these place no constraints on the \emph{dd} model parameter space shown). We also show regions that are above the upper limits from gamma-ray observations of dwarf galaxies (purple). In both models, the masses of the heavy scalar and charged Higgs are fixed to be $300$ GeV. The pseudoscalar mass is 160 GeV.
\label{fig:GCEplots}}
\end{center}
\end{figure*}

We will represent the regions with the correct relic density in black in our summary plots of Figs.~\ref{fig:scan1} and~\ref{fig:GCEplots}.  Regions where the annihilation rate is above the 95 \% CL upper limits from gamma-ray observations of dwarf spheroidals will be represented in purple.

\vspace{.5cm}
{\bf Direct Detection.}
The dominant contribution to the SI scattering rate of $\chi$ off of nuclei is from exchange of the CP-even scalars $h$, $H$. The SI $\chi$ scattering cross section per nucleon is
\begin{equation}
\sigma_0^\text{SI, per nucleon} = \frac{4 \mu^2_{\chi,n} }{\pi} \Big[ \frac{Z f_\text{prot} + (A-Z) f_\text{neut}}{A} \Big]^2
~,
\label{SIsigma}
\end{equation}
where $A$ and $Z$ are the atomic mass and atomic number, respectively, of the target nucleus, and $\mu_{\chi,n}$ is the reduced mass of the WIMP-nucleon system. $f_n$ (where the nucleon $n$ is a proton or neutron) is the effective WIMP coupling to nucleons, which can be written in terms of the quark couplings as
\begin{equation}
f_n \equiv m_n \bigg[ \sum\limits_{q= u,d,s} \frac{a_q}{m_q} f_{Tq}^{(n)} + \frac{2}{27} f_{TG}^{(n)} \sum\limits_{q = c,b,t} \frac{a_q}{m_q}  \bigg]
~,
\label{SInucleon}
\end{equation}
where $a_q/m_q$ is defined to be
\begin{align}
\frac{a_q}{m_q} &\equiv \frac{1}{v} \left[ - \frac{\lambda_{\chi h}}{m_h^2} + \frac{\lambda_{\chi H} q _{\beta H}}{m_H^2} \right]
\nonumber \\
q_{\beta H} &\equiv \begin{cases} \cot{\beta} , & \text{if }q=\text{up-type}  \\ -\tan{\beta}, & \text{if } q=\text{down-type .}  \end{cases}
\label{SIquark}
\end{align}
In Eq. (\ref{SInucleon}), we take the quark mass fractions to have the values $f_{T_u}^\text{(prot)} = 0.02$, $f_{T_d}^\text{(prot)} = 0.026$, $f_{T_u}^\text{(neut)} = 0.014$, $f_{T_d}^\text{(neut)} = 0.036$, $f_{T_s}^\text{(prot)} = f_{T_s}^\text{(neut)} = 0.043$. By definition, $f_{TG}^{(n)} = 1 - \sum\limits_{q=u,d,s} f_{T_q}^{(n)}$~\cite{Junnarkar:2013ac,Crivellin:2013ipa}. The effective WIMP-Higgs Yukawa couplings in Eq.~(\ref{SIquark}) are the same ones given explicitly 
in Eqs.~(\ref{duHiggs})-(\ref{uuHiggs}).
Throughout, we demand that the rate given in Eq.~(\ref{SIsigma}) is below the SI constraints from LUX  \cite{Akerib:2013tjd}. We will represent this constraint in red in our summary plots of Figs.~\ref{fig:scan1} and~\ref{fig:GCEplots}.

As we will see later in Sec.~\ref{discussion}, the sign of $\tan{\theta}$ can have important effects on the direct detection rate. Although it will only slightly alter the DM annihilation rate throughout its thermal history, a negative $\tan{\theta}$ can allow for a suppressed SI scattering rate as seen by LUX. To illustrate this effect, we show in Fig.~\ref{SIdudd} the SI nucleon scattering rate (normalized by the LUX limit) for the \emph{du} and \emph{dd} models as a function of $\tan{\theta}$ and for various representative choices of $y$, $M_S$, and $M_D$. 

In the left plot of Fig.~\ref{SIdudd}, we show the normalized rate for the \emph{du} model.  For $\tan{\theta}<0$, Eq.~(\ref{duHiggs}) and Eq.~(\ref{SIquark}) imply that the effective light and heavy Higgs couplings with nucleons can be comparable and opposite in sign. As can be seen in Fig.~\ref{SIdudd}, this destructive interference between $h$ and $H$ exchange is a generic feature of the \emph{du} model for $\tan{\theta}<0$, however the exact point of maximum cancellation depends on the specific choice of $y$, $M_S$, $M_D$, and $\tan{\beta}$.  Furthermore, independent of the chosen benchmark of couplings, for large $\tan\beta$, there is always a suppression at $\tan{\theta} = y_2/y_1 = 0$, which can be understood as the point of suppressed mixing since both mixing terms $y_1 v_1 = y v_d \cos \theta$ and $y_2 v_2 = y v_u \sin \theta$ are small here.

In the right plot of Fig.~\ref{SIdudd}, we show the normalized SI scattering rate for the \emph{dd} model. In this case, the rate can also be suppressed for  $\tan{\theta}<0$. This is because both couplings $\lambda_{\chi h}$ and $\lambda_{\chi H}$ vanish at a blind spot of the theory when $m_\chi + M_D \sin{2 \theta} = 0$. As can be seen in the figure, the position of the blind spot does not depend much on the values of $y$ and $m_H$. For the $M_S$ and $M_D$ parameters in  Fig.~\ref{SIdudd}, this occurs at $\tan \theta \approx - 0.15, -6.51$.

We next consider elastic spin-dependent (SD) scattering of $\chi$ with nuclei via $Z$ exchange, requiring that the DM-neutron cross section\footnote{The constraints on SD scattering with neutrons is generally stronger than that coming from protons.} is consistent with the recast constraints from LUX \cite{Savage:2015xta}. The SD scattering of $\chi$ per neutron is
\begin{equation}
\sigma_0^\text{SD, per neutron} = \frac{12 \mu^2_{\chi, \text{neut}}}{\pi} \Big( \sum\limits_{q=u,d,s} a_q  \Delta_q^{(\text{neut})} \Big)^2
~,
\label{SDsigma}
\end{equation}
where  $a_q$ is the effective coupling with quarks,
\begin{align}
a_q &\equiv \frac{g_{\chi Z} g_{q a}}{m_Z^2}
\nonumber \\
g_{q a}  &\equiv \mp \frac{e}{4} ( t_w + t_w^{-1})~,\text{ if $q=$ up/down-type}
~.
\end{align}
In Eq.~(\ref{SDsigma}), we take the quark spin fractions to be $\Delta_u^{(\text{neut})} = -0.42$, $\Delta_d^{(\text{neut})} = 0.85$, and $\Delta_s^{(\text{neut})} = -0.08$~\cite{Cheng:2012qr}. 
$g_{\chi Z}$ is the coupling of DM with the $Z$ boson and is given in Eq.~(\ref{GaugeCouplings}). Here, $e$ is the electric charge of the electron ($e>0$), and $t_w$ is the tangent of the Weinberg angle. 

The constraints on the SD scattering off of neutrons from LUX become more important as the doublet mass $M_D$ is decreased and the doublet fraction is correspondingly enhanced (of course, if $\chi$ is purely doublet, then $g_{\chi Z}$ vanishes completely as seen in Eq.~(\ref{GaugeCouplings})). We will represent this constraint in orange in our summary plots of Figs.~\ref{fig:scan1} and~\ref{fig:GCEplots}.

\vspace{.5cm}
{\bf Invisible Decays.}
The constraints from the invisible widths of the SM Higgs $h$ and $Z$ are relevant whenever $m_\chi \lesssim m_h/2,~m_Z/2$, respectively. The invisible branching fraction of the Higgs is constrained to satisfy BF$(h \to \chi \chi) \lesssim 0.19$ at 95\% CL, which comes from a global fit to the visible Higgs channels in which the visible Higgs couplings are fixed to their SM values \cite{Belanger:2013xza}. Assuming that $\Gamma(h \to \text{SM}) = \Gamma^\text{SM}(h \to \text{SM})$, it follows that the constraint on the invisible width of the Higgs is approximately $\Gamma ( h \to \chi \chi ) \lesssim 0.99$ MeV. From the Lagrangian of Eq.~(\ref{HiggsL}), the Higgs width into a pair of Majorana $\chi$'s is found to be
\begin{equation}
\Gamma (h \to \chi \chi) = \frac{\lambda_{\chi h}^2}{4 \pi} m_h \left( 1 - \frac{4m_\chi^2}{m_h^2} \right)^{3/2}
~.
\end{equation}
Similarly, electroweak precision measurements at LEP constrain $\Gamma( Z \to \chi \chi ) \lesssim 2$ MeV  \cite{ALEPH:2005ab}. For $m_\chi \lesssim m_Z/2$, the Z width into a pair of $\chi$'s is
\begin{equation}
\Gamma (Z \to \chi \chi) = \frac{g_{\chi Z}^2}{6 \pi} m_Z \left( 1 - \frac{4m_\chi^2}{m_Z^2} \right)^{3/2}
~.
\end{equation}
We will represent the constraint from the Higgs ($Z$) invisible width in gray (brown) in our summary plot of Fig.~\ref{fig:GCEplots}.

\vspace{.5cm}
{\bf Direct searches for new particles.}
The model contains additional (possibly light) new particles that can be looked for directly at the LHC: two neutral ($H,A$) and one charged Higgs boson ($H^\pm$) and two neutral ($\chi_{2,3}$) and one charged ($E$) fermion, in addition to the DM candidate, $\chi \equiv \chi_1$. In Sec.~\ref{sec:Presentbounds}, we have already discussed the LHC bounds on neutral heavy Higgs bosons. Here we comment on the constraints on charged Higgs bosons, as well as on new fermions.

The most relevant LHC charged Higgs searches are for the process $pp\to t H^\pm, t b H^\pm$, with subsequent $H^\pm \to\tau\nu$~\cite{ATLASchargedH,CMS:2014cdp} or $H^\pm \to t b$~\cite{CMS:2014pea}. These searches, performed with 8 TeV data, probe charged Higgs bosons in the multi-hundred GeV range, but only for very large values of $\tan\beta$ ($\gtrsim 30-40$), for which the coupling $b_R t_L H^\pm$ entering the charged Higgs production is enhanced. 
In addition to direct searches, flavor transitions such as $b \to s \gamma$ can set interesting (indirect) bounds on the charged Higgs mass: in a Type-II 2HDM, charged Higgs bosons cannot be lighter than around (300-400) GeV~\cite{Hermann:2012fc}. In the following, we will always fix the charged Higgs mass to 300 GeV and  $\tan\beta\lesssim 10$ in such a way as to avoid constraints from flavor transitions and direct collider searches for heavy Higgs bosons.

LHC searches for electroweak Drell-Yan production can set interesting bounds for the new fermions arising in our model. In particular, the model contains one charged and two neutral fermions, in addition to the DM. These fermions are produced either in pairs through a $Z$ boson ($pp\to \chi_i\chi_i$, where $\chi_i= \chi_{2,3},\,E$), or in associated neutral-charged production, with the exchange of a $W$ boson ($pp\to \chi_{2,3} E$). Generically, the latter production mode has the most relevant LHC constraints. LHC searches for supersymmetric Wino associated production giving a $3\ell\,+$ (or $2\ell\,+$) MET signature can already set bounds on the Wino mass at around 400 GeV for massless lightest supersymmetric particles (LSPs)~\cite{Khachatryan:2014qwa,Aad:2014vma,Aad:2014nua}. In Sec.~\ref{GCE}, we will discuss how this bound can be interpreted in our model.


\subsection{Discussion on DM above 100 GeV}
\label{discussion}

Among all of the interactions of our DM candidate $\chi$, the coupling to the pseudoscalar $A$ is key in opening up viable parameter space.  Annihilation through an $s$-channel pseudoscalar is $s$-wave, and if it happens not too far from the pseudoscalar resonance,
\begin{equation}
	1-4 m_\chi^2/m_A^2 \lesssim 0.5,
\end{equation}
then it is possible to obtain $\Omega h^2 \sim 0.12$ even for a large singlet fraction.  And unlike the interactions of $\chi$ with $h$ and $H$, $\chi$ scattering off of quarks via exchange of the pseudoscalar $A$ is spin-dependent (SD) and further kinematically suppressed by four powers of the momentum transfer, $q^4$, where $q$ is typically of order 100 MeV. 

In exploring the parameter space of the model, we have found that it is difficult to find thermal DM candidates very far away from pseudoscalar resonances. When the relic density is not governed by resonances (or coannihilations) to any significant degree, there must be large mixing in the gauge eigenstate makeup of $\chi$ in order to obtain the correct abundance.  Since this mixing hinges on the Yukawa interactions (see Eq.~(\ref{MixingAngles})), this will also increase the DM scattering rate off of quarks. Such regions are generally ruled out by LUX, except near special blind spots, as discussed above.

We present benchmarks for two of the models of Table~\ref{modelnames} in Fig.~\ref{fig:scan1}. As previously mentioned, the \emph{du} and \emph{ud} models can be related to each other by replacing $\tan{\theta}$ with its inverse, and the \emph{uu} model in the large $\tan{\beta}$ limit reduces to only coupling the SM Higgs to the dark sector. We therefore only explore the parameter space for the models \emph{du} and \emph{dd}.   

In this section, we focus on $m_\chi \gtrsim 100$ GeV, which also easily avoids constraints from the invisible width of the $Z$ or $h$. Prospects for lighter DM ($m_\chi \lesssim 100$ GeV) will be presented below in Sec.~\ref{GCE}. We also choose to work with sufficiently light $\chi$, such that annihilations to final states including one or more heavy Higgs are negligible. This therefore favors scalars of mass around a few hundred GeV in order for a pseudoscalar resonance to be relevant, and so we fix $m_A = m_H = m_{H^\pm} = 300$ GeV in order to simplify the scan of the parameter space. We note the model would work just as well for heavier scalars and correspondingly heavier DM. 

In. Fig.~\ref{fig:scan1}, we show constraints in the $(M_D, M_S)$ plane for representative choices of $y$, $\tan{\theta}$, and $\tan{\beta}$ that give viable parameter space for thermal relic DM. Our choice of parameters satisfies the constraints from direct searches for heavy Higgs particles in the $b \tau \tau $ and $bbb$ final states, as discussed in Section~\ref{sec:collider}. Furthermore, since the pseudoscalar-DM coupling $\lambda_{\chi A} \lesssim \mathcal{O}(0.1)$ in the full $M_D-M_S$ plane we present, mono-$b$ searches are much less sensitive to this model.

In both \emph{du} and \emph{dd} models, we clearly see the least constrained region for thermal relic DM is where $m_\chi \lesssim 150$ GeV ($m_\chi$ at or slightly above 150 GeV is in slight tension with gamma-ray observations of dwarf galaxies). The thermal relic line extends to larger $M_D$ (larger singlet fraction) as the DM mass approaches the resonant region, which is centered  slightly below $m_\chi = m_A/2$. This can be understood as thermal broadening of the resonance near freeze-out. Another feature in the thermal relic line can emerge if $\tan{\beta}$ is not too large: in the right frame of Fig.~\ref{fig:scan1}, for $m_\chi \gtrsim 175$ GeV, there can be dominant annihilations to top quarks through $s$-channel exchange of a light or heavy Higgs. 

Although the pseudoscalar resonance region will remain viable for many different parameters, the proximity or tuning of $m_\chi$ to $m_A/2$ depends on the choice of $\tan \theta$ or $y$.  The thermal relic line in the figures will shift  to 
larger values of $M_D$ for a fixed $M_S$ for large $|\tan{\theta}|$.  (The relic abundance is only slightly affected by the sign of $\tan \theta$.) This is because, in both \emph{du} and \emph{dd} models, for large $|\tan{\theta}|$,  $|y_2 v_2 |\gg |y_1 v_1|$, which then implies (using Eq.~(\ref{MixingAngles})) $|N_1^\chi| \gg |N_2^\chi|$. Hence, $g_{\chi Z}$ is largely unsuppressed, as seen by Eq.~(\ref{GaugeCouplings}), and to compensate, the overall singlet fraction $N_s^\chi$ must be increased by slightly decoupling $M_D$. Similarly, the doublet fraction of $\chi$ is proportional to $y$, and so increasing $y$ in either of the scans of Fig.~\ref{fig:scan1} will generically shift the thermal relic line to larger values of $M_D$ for a given value of $M_S$. 

Direct detection constraints are most relevant at lower $M_D$, where the DM singlet fraction is lower (see the red region of Fig.~\ref{fig:scan1}). Fig.~\ref{fig:scan1} also illustrates various  blind-spots in the SI direct detection rate. In the \emph{du} model, for $M_D \sim 100-200$ GeV, the choice of $\tan{\theta} < 0$ suppresses nucleon scattering. In particular, from Eq.~({\ref{duHiggs}}), in the large $\tan{\beta}$ limit the relative strength of the two different CP-even Higgs couplings is $\lambda_{\chi h}^{du} / \lambda_{\chi H}^{du} \sim (m_\chi / M_D) \tan{\theta}$. Then, from Eq.~(\ref{SIquark}), when $\tan{\theta}=-1$ the couplings of nucleons to $h$ and $H$ partially cancel, explaining the feature in the drop off in scattering rate for $M_D \sim 200$ GeV and $M_S \sim 100$ GeV.
In the \emph{dd} model, shown in the right frame of Fig.~\ref{fig:scan1}, $\chi$-nucleon SI scattering is also near a blind-spot of the model where both $\lambda_{\chi h},\lambda_{\chi H}$ are suppressed. Here, fixing $\tan{\theta} = -10$, $\lambda_{\chi h} \approx \lambda_{\chi H} \approx 0$ approximately when $M_D/m_\chi \approx 5$.

The SD constraints (the orange region in Fig.~\ref{fig:scan1}) do not rule out much of the viable thermal relic parameter space that SI constraints do not already exclude. The importance of considering SD scattering via $Z$ exchange is still illustrated in the $\emph{du}$ model, where spin-dependent limits are more powerful than spin-independent limits for $M_D$ values of $\lesssim 200$ GeV.  The SD limits do not depend on Higgs couplings and are able to constrain the parts of the parameter space close to where $h$ and $H$ exchange interfere and suppress the SI scattering rate.

Finally, as already introduced in the previous subsection, additional constraints might come from the LHC direct search of electroweak Drell-Yan production. However, in the regime with not too light DM, $m_\chi\gtrsim(100-150)$ GeV,  there are no bounds even for $M_D$ of around 200 GeV~\cite{Khachatryan:2014qwa,Aad:2014vma}.

\subsection{Light DM and the GCE}
\label{GCE}

In this section, we investigate the viability of the models to describe DM with mass below $\sim 100$ GeV. Although much of the physics near the pseudoscalar resonance is similar to that of the previous section, we additionally require that the model could provide a reasonable fit of the Galactic Center Excess. For model building in a similar direction that can additionally describe the 3.55 keV X-ray line, see~\cite{Berlin:2015sia}.

One simplified model that has received much attention in its ability to describe this signal is just that of Eq.~(\ref{eq:Lsimple}). Since the annihilation is $s$-wave, the rate can still be large today. Moreover, a relatively light pseudoscalar is favored in UV-complete realizations to get around numerous constraints~\cite{Cheung:2014lqa,Cahill-Rowley:2014ora,Ipek:2014gua}. The same is true in our models, since a pseudoscalar of around 100-200 GeV will be needed for the DM to annihilate near-resonance. 
As discussed in detail in Sec.~\ref{2HDM}, we therefore implement the freedom in a general 2HDM to have a sizable splitting between the pseudoscalar mass and the heavy/charged Higgs mass and fix $m_A = 160$ GeV and $m_H = m_{H^{\pm}} = 300$ GeV. This allows $\chi$ to be relatively close to a pseudoscalar resonance, while the other scalars are heavy to evade direct detection and other constraints.

We now add the following criteria to the list of demands enumerated in Sec.~\ref{generalconstraints} for $\chi$ to be considered a realistic DM candidate \emph{consistent with the GCE}:

\begin{enumerate}
\item Since annihilations through a light pseudoscalar proceed dominantly to final state bottom quarks, we restrict the mass of $\chi$ to lie in the range $50\text{ GeV}~\leq~m_\chi~\leq~70\text{ GeV}$ in order to fit the spectral shape of the GCE spectrum (this is represented in blue in Fig.~\ref{fig:GCEplots}).\footnote{The exact mass range that is preferred is dependent on systematics. For annihilations to $b\bar{b}$, the spectral shape of the observed emission has been found to be well fit by DM of mass as low as 30 GeV and as high as 70 GeV~\cite{Daylan:2014rsa,Abazajian:2014fta,Calore:2014xka,Calore:2014nla}. As a benchmark, we choose the upper half of this range since it is more viable for the model at hand.}
\item In order to ensure the approximate normalization for the GCE signal, the annihilation rate must satisfy $0.5\times10^{-26} \text{ cm}^3/\text{s} \leq \lim_{v \to 0} \langle \sigma v \rangle \leq 4  \times 10^{-26} \text{ cm}^3/\text{s}$~\cite{Calore:2014nla} (this is represented in green in Fig.~\ref{fig:GCEplots}).
\end{enumerate}
Limits on DM annihilation from AMS-02 observations of the anti-proton fraction~\cite{Giesen:2015ufa,Lin:2015taa} are very similar to those from the dwarf galaxies and have additional astrophysical uncertainties, so  for simplicity  we do not include them.

In Fig.~\ref{fig:GCEplots} we show benchmark scenarios for $m_\chi \lesssim 100$ GeV, highlighting the regions that fit the GCE. We again consider the  \emph{du} and \emph{dd} models and scan over  the $(M_D,M_S)$ plane.
We make a similar choice for $y$, $\tan{\theta}$, and $\tan \beta$ as in Sec.~\ref{discussion}. For both models, the qualitative behavior of the direct detection constraints is similar to the previous section. Again, since the pseudoscalar-DM coupling $\lambda_{\chi A} \lesssim \mathcal{O}(0.1)$, mono-$b$ searches would have very little sensitivity to the relevant parameter space.

For both  models in Fig.~\ref{fig:GCEplots}, the GCE excess can be explained while avoiding direct detection limits if $\chi$ is mostly singlet-like and near a pseudoscalar resonance. Constraints from gamma-ray observations of dwarf spheroidals may also be avoided if $\chi$ is slightly below the pseudoscalar resonance, due to thermal averaging of the annihilation cross section. We also see the  effects of annihilation near the $Z, h$ poles (the latter only for the \emph{du} model), which is visible in the thermal relic contour at around $M_S \sim 45, 60$ GeV, respectively. This effect is not present for  $\lim_{v \to 0} \langle \sigma v \rangle$; for Majorana fermions annihilating through an $s$-channel vector mediator (or Dirac fermions with only axial couplings) there is no resonant enhancement in the $s$-wave contribution to $\sigma v$~\cite{Jungman:1995df} and annihilation through an $s$-channel scalar mediator is $p$-wave for fermionic DM. 

The favored region for the \emph{du} model is for $M_D \sim 300\,{\rm{GeV}}-1$ TeV, and $m_\chi \sim M_S = 60-70 \text{ GeV}$. Due to the relatively smaller mixing induced in the \emph{dd} model, the appropriate GCE parameter space requires a smaller $M_D \sim (200 - 500)$ GeV. We also emphasize that, although we have presented only two benchmark scenarios here, these choices of models and parameters of the DM sector are not particularly special or highly fine-tuned. In particular, all models with a pseudoscalar that is not too heavy, $\tan\theta<0$, and not too small $\tan\beta$ could give a good fit of the GCE. Of course, pseudoscalars even lighter than 160 GeV would be suitable to obtain a large enough annihilation rate. For example, a pseudoscalar mass around 100 GeV is even better suited for the GCE. For lighter $m_A \lesssim 100$ GeV, then there are currently no LHC direct searches for pseudoscalars (although this may be possible in the future \cite{Kozaczuk:2015bea}) and only very weak constraints from heavy Higgs searches at LEP \cite{Schael:2006cr} are applicable. However, such very light pseudoscalars are more difficult to be achieved within our 2HDM scalar sector (see Appendix~\ref{app:Higgsspectrum} for more details). 

We finally comment on the additional constraints we have to consider for these models with light DM, and consequently with new relatively light electrically charged degrees of freedom ($E$ in Eq.~(\ref{HiggsL}) and Eq.~(\ref{eq:gaugeLagrangian})). We first note that the constraint from  the invisible width of the SM $h$ and $Z$ are not strong and can exclude only a small region of parameter space in the \emph{du} model at light $M_D$, which is already excluded by LUX constraints.

Additionally, having new light and $SU(2)$ charged degrees of freedom can introduce corrections to electroweak precision observables. However, we have checked that the contributions to the $T$ parameter from loops of the new fermions is negligible (at most at the level of $\Delta T \sim 10^{-4}-10^{-3}$ in the region of parameter space favored by the GCE)~\cite{Baak:2014ora}.

Finally, constraints on the parameter space come from LHC direct searches for Drell-Yan production of electroweak particles. For the models shown in Fig~\ref{fig:GCEplots}, the electroweak spectrum needed to fit the GCE contains several new light fermions in addition to DM. In particular, for the \emph{dd} model, we have two additional neutral fermions $\chi_{2,3}$ and one charged fermion $E$,  all with mass close to $M_D \sim(200-500)$ GeV range and with splittings smaller than a few GeV amongst the states.
Constraints from LHC  searches of neutral-charged Wino associated production, resulting in a $3\ell\,+$ (or $2\ell\,+$) MET signature, are the most important to constrain our scenario. In particular, combining the $2\ell$ and $3\ell$ searches, the ATLAS collaboration sets a bound at $m_{\rm{Wino}}\sim 400$ GeV, under the assumption of $100\%$ branching ratio $\tilde W^\pm\tilde W^0 \to W^\pm Z+ \missET$ and massless LSP~\cite{Aad:2014vma}. This corresponds to an exclusion on $\sigma(pp\to\tilde W^\pm\tilde W^0 \to W^\pm Z+ \missET)\gtrsim 30$ fb. In the following, we discuss how to interpret this constraint in terms of our model.

As discussed in Sec.~\ref{themodel}, the fermion content of our model resembles the one of the MSSM with a Bino-like LSP and Higgsino-like NLSPs. However, in our model the heaviest fermions $\chi_2$ and $\chi_3$ have a sizable branching ratio $\chi_{2,3}\to A\chi_1$, as long as it is kinematically accessible ($M_D\gtrsim 220$ GeV for the benchmarks in Fig.~\ref{fig:GCEplots}). Then it is easy to check that, in the \emph{dd} model, the cross section for $pp\to \chi_{2,3}E\to WZ\chi_1\chi_1$ is always smaller than the excluded cross section (30 fb) in the entire region of parameter space for $M_D\gtrsim 220$ GeV. Below 220 GeV, the decay into a pseudoscalar is not accessible and the branching ratio for the decay $\chi_{2,3}\to h \chi_1$  is not large enough to suppress sufficiently the $\chi_{2,3}E\to WZ\chi_1\chi_1$ channel. Therefore, in the \emph{dd} model, the region $M_D\lesssim 220$ GeV of Fig.~\ref{fig:GCEplots} has already been  probed by the LHC direct searches of Drell-Yan production of electroweak particles. On the other hand, the region of parameter space favored by the GCE  in the \emph{du} model has not been probed by these LHC searches yet, since the additional fermions are heavier ($M_D\gtrsim 300$ GeV).

\subsection{Future tests of the model. \label{sec:futuretest}}

Upcoming spin-independent direct detection results can probe much of the parameter space for our model. For the $du$ model, which has a larger SI rate compared to the $dd$ model, future LUX data can test the entire GC excess region and much of the region for heavy dark matter. The next generation of ton-scale direct detection experiments, such as LZ~\cite{LZ} and XENON1T~\cite{XENON1T}, will be vastly more constraining. The LZ experiment can cover the entirety of the parameter space shown for the $du$ model in Figs.~\ref{fig:scan1}-\ref{fig:GCEplots}, while for the $dd$ model a significant portion of the parameter space can be reached, including the entire GC excess region shown in Fig.~\ref{fig:GCEplots}.

Our model can be further probed at LHC Run II by the search of the various light degrees of freedom. In particular, as shown in Fig.~\ref{fig:Aparamspace}, pseudoscalar searches (with the pseudoscalar decaying either to taus or invisibly) will be able to test almost entirely the region of parameter space able to predict a thermal DM candidate, if both the DM and pseudoscalar are relatively light ( $m_{\textrm{DM}} < m_A/2$ and $m_A$ below $\sim 200$ GeV). Furthermore, at the Run II of the LHC, with 300 fb$^{-1}$ data, searches for Drell-Yan production of electroweak particles will be able to probe Wino masses as high as $\sim 840$ GeV~\cite{EWinoProjections}, under the assumption of $100\%$ decay for $\tilde W^\pm\to W + \missET$ and for $\tilde W^0\to Z + \missET$. This can be translated into a bound on $M_D$ at the level of $\sim 300$ GeV in the \emph{dd} model for the GCE region\footnote{Note that for $M_D$ above $\sim 350$ GeV, a new decay mode for $E$ becomes available $E\to H^\pm \chi_1$, suppressing even more the cross section for the process $pp\to \chi_{2,3}E\to WZ+ \missET$.}.

Finally, our model also predicts additional signatures that can be searched for at the LHC.  As discussed above, Drell-Yan production can dominantly lead to $pp\to \chi_{2,3}E\to WA\chi_1\chi_1$, with $A \to \bar b b, \tau \tau$, or even $\chi_1 \chi_1$. Similar signals have $A$ replaced by $h$. In particular, for $M_D\leq 300$ GeV, the new scalars will have sizable branching ratios into the $\chi_i$ and $E$ fermions: the charged Higgs can decay to DM and to a charged fermion $E$, $H^\pm\to E \chi_1$, producing a $W+\missET$ signature; the heavy scalar can decay to DM and to an additional neutral fermion $H\to \chi_1\chi_{2,3}$, with $\chi_{2,3}$ possibly decaying to $\chi_1$ plus a $Z, h$, or $A$. It will be interesting to investigate the potential of the $pp\to t b H^\pm,\, H^\pm\to W +\missET$ and $pp\to t t \,(bb)\, H,\, H\to \chi_1\chi_{2,3}\to 2\ell\,(2b,2\tau)+\missET$ channels in probing our model.

\section{Conclusions \label{sec:conc}}

Viable theories of weak scale dark matter are under increasing pressure from the unyielding progress of current direct detection and collider experiments. Simplified models of dark matter that introduce preferred pseudoscalar interactions with quarks explain the null results of LUX, while still motivating interesting collider searches in the form of mono-$b$ or $t \bar t + \missET$ processes. If the pseudoscalar is introduced in the context of a general 2HDM, we have shown how the freedom in the scalar potential allow the pseudoscalar to be slightly decoupled, $m_A < m_H \sim m_{H^\pm}$. Motivated by this framework, we have presented current and projected limits from missing energy and  heavy Higgs searches at the LHC at energies of 8 and 14 TeV.

We have also introduced a concrete realization of a model that couples a Type-II 2HDM to a fermionic dark sector consisting of a singlet and a pair of $SU(2)$ doublets. The lightest $\mathbb{Z}_2$ odd particle, $\chi$, is a dark matter candidate that possesses couplings to the scalars after mixing between the singlet and doublet.  This model has analogues in the MSSM with bino-higgsino DM, and singlino-higgsino DM in the NMSSM, but we allow arbitrary couplings and scalar spectra.  To simplify the presentation, we consider a discrete subset of couplings of the $SU(2)$ doublets to the two Higgs doublets. Considering direct detection, $h$ and $Z$ invisible decay, and other collider constraints, we identify regions of parameter space with thermal relic DM below a few hundred GeV. In all of the parameter space that we have presented, thermal relic $\chi$ remains unconstrained if its mass is in proximity to a pseudoscalar resonance.

For  $m_\chi \lesssim 100$ GeV, this model can describe the excess gamma-rays coming from the GC as measured from the \Fermi satellite. A good fit to the gamma-ray data is given by DM annihilation to $b$-quarks with DM mass in the range 50-70 GeV. For these masses, we find that the preferred scalar spectra has a light pseudoscalar of around 160 GeV and heavier scalars at around 300 GeV. Future direct detection data from LZ will be able to thoroughly test these scenarios. There are also numerous LHC tests of the model, in the context of the GC excess.  Precision Higgs coupling measurements can constrain the 2HDM that give rise to such light pseudoscalars. LHC searches for jets plus MET or the (in)visible decays of new heavy scalars can cover much of the parameter space at the 14 TeV LHC. Consistency with the GCE generically also predicts new electrically charged states in the mass range of $200-1000$ GeV.   At Run 2, direct production of these new states can be searched for in the $3\ell+\missET$ channel.  In addition, there are additional new collider signals with MET, heavy flavor, and gauge or Higgs bosons that result from the model. These new signatures may be valuable to study in future LHC data, independently of the specific model we have presented.

\begin{acknowledgments}
	We thank Dan Hooper, Maxim Pospelov and Carlos Wagner for useful discussions.
	This work was supported in part by the Kavli Institute for Cosmological Physics at the University of Chicago through grant NSF PHY-1125897 and an endowment from the Kavli Foundation and its founder Fred Kavli. SG and TL would like to thank the Center for Future High Energy Physics (CFHEP) in Beijing for hospitality and partial support. Research at Perimeter Institute is supported by the Government of Canada through Industry Canada and by the Province of Ontario through the Ministry of Economic Development $\&$ Innovation. Part of this work was completed at the Aspen Center for Physics, which operates under the NSF Grant 1066293.
\end{acknowledgments}

\onecolumngrid
\appendix

\section{Higgs spectrum \label{app:Higgsspectrum} }

In this Appendix, we comment on a feasible configuration of quartic couplings that is able to produce the Higgs spectrum presented in Sec.~\ref{GCE}. In particular, we want to achieve a sizable splitting between the pseudoscalar mass $m_A$ and the heavy neutral and charged Higgs, as well as an approximate alignment limit $\cos(\alpha-\beta)\sim 0$, for which the 125 GeV Higgs ($h$) has SM-like properties.

To achieve a large splitting between the pseudoscalar and the charged Higgs, we need a sizable $\lambda_5-\lambda_4$ (see Eq. (\ref{eq:mCminusmA})). Additionally, the value of $\lambda_2$ is fairly constrained by the requirement of having the lightest scalar with mass at around 125 GeV (see Eq. (\ref{eq:mh})). Furthermore, in the regime of sizable $\tan\beta$, to get the alignment limit, we need to have the off-diagonal term for the mass matrix in the $(h_u,h_d)$ basis much smaller than the diagonal terms. This translates into the condition $\lambda_3+\lambda_4\sim m_A^2/v^2$. Finally, our quartic couplings have to satisfy the conditions for a potential bounded from below, and, in particular, $\lambda_3\geq -\lambda_4+|\lambda_5|-\sqrt{\lambda_1\lambda_2}$~\cite{Branco:2011iw}. Putting together these conditions, we learn that also the value of $\lambda_1$ should be sizable.

A possible benchmark that produces the Higgs spectrum presented in Sec.~\ref{GCE} is given by
\beq
\lambda_1=2.2,\,\lambda_2=0.24,\,\lambda_3=1.4,\,\lambda_4=-1.05,\,\lambda_5=1.05,\, m_A=160\,{\rm{GeV}},
\eeq
that leads to $m_h\sim 126$ GeV, $m_H\sim m_{H^\pm}\sim 300$ GeV, and relatively close to the alignment limit ($\cos(\alpha-\beta)\sim 0.1$).

From this discussion, we also learn that it will be difficult to have a much larger splitting between the pseudoscalar and the charged/heavy Higgs, if we want to maintain a stable potential and not too huge values of the quartic couplings.

This benchmark scenario should be compared to the quartic coupling configuration of the MSSM:
\begin{eqnarray}
&&\lambda_1^{\rm{MSSM}}=\lambda_2^{\rm{MSSM}}=\frac{g^2+g_1^2}{4}\sim ~ 0.14~,  ~~\,\lambda_3^{\rm{MSSM}}=\frac{g^2-g_1^2}{4}\sim 0.08~,~~\, \nonumber \\
&&\lambda_4^{\rm{MSSM}}=-\frac{g^2}{2}\sim -0.21~,~~\,\lambda_5^{\rm{MSSM}}=0\,,
\end{eqnarray}
where $g$ and $g_1$ are the $SU(2)$ and $U(1)$ gauge coupling, respectively.

\section{Higgs couplings in singlet-doublet model \label{app:higgscouplings} }

We will parametrize the Higgs doublets as
\begin{equation}
\Phi_{1,2} = \frac{1}{\sqrt{2}} \begin{pmatrix} \sqrt{2} N_{1,2}^+ H^+ \\ v_{1,2} + N^h_{1,2}h + N^H_{1,2}H + i N^A_{1,2}A  \end{pmatrix} ~,
\label{HiggsParam}
\end{equation}
where, in the alignment limit, the coefficients $N_{d,u}^{h,H,A,+}$ are
\begin{align}
N_d^h &= \cos{\beta} ~,~ N_u^h = \sin{\beta} ~,~ N_d^H = \sin{\beta} ~,~ N_u^H = -\cos{\beta}
\nonumber \\
\nonumber \\
N_d^A &= - \sin{\beta} ~,~ N_u^A = \cos{\beta} ~,~ N_d^+ = -\sin{\beta} ~,~ N_u^+ = \cos{\beta} 
~.
\end{align}
The Higgs couplings of Eq.~(\ref{HiggsL}) in terms of mixing angles are
\begin{align}
\lambda_{\chi h} &\equiv - \frac{1}{\sqrt{2}}N^\chi_S \left( y_1 N^\chi_1 N^{h}_1 + y_2 N^\chi_2 N^{h}_2 \right)
\nonumber \\
\lambda_{\chi H} &\equiv - \frac{1}{\sqrt{2}}N^\chi_S \left( y_1 N^\chi_1 N^{H}_1 + y_2 N^\chi_2 N^{H}_2 \right)
\nonumber \\
\lambda_{\chi A} &\equiv \frac{1}{\sqrt{2}}N^\chi_S \left( y_1 N^\chi_1 N^A_1 - y_2 N^\chi_2 N^A_2 \right)
\nonumber \\
\lambda_{+ s} &\equiv \frac{1}{2} N_S^\chi  \left( y_1 N_1^+ + y_2 N_2^+ \right)
\nonumber \\
\lambda_{+ p} &\equiv -\frac{1}{2} N_S^\chi \left( y_1 N_1^+ - y_2 N_2^+ \right)
~.
\label{HiggsCouplings}
\end{align}

Analytic results are obtained by taking partial derivatives with respect to $v_1$ or $v_2$ of the characteristic equation Det$\left(M_\text{neutral}-m_\chi \right)=0$ and then solving for the couplings $\lambda_{\chi 1}$ and $\lambda_{\chi 2}$ defined by 
\begin{equation}
\lambda_{\chi 1} \equiv -\frac{1}{2}\frac{\partial m_\chi} {\partial v_1} ~,~ \lambda_{\chi 2} \equiv -\frac{1}{2}\frac{\partial m_\chi}{\partial v_2}
~.
\end{equation}
With these unphysical couplings in hand, the physical ones to $h$, $H$, $A$ can be obtained,
\begin{align}
\lambda_{\chi h} &= N_1^h \lambda_{\chi 1} + N_2^h \lambda_{\chi 2}
\nonumber \\
\lambda_{\chi H} &= N_1^H \lambda_{\chi 1} + N_2^H \lambda_{\chi 2}
\nonumber \\
\lambda_{\chi A} &= -N_1^A \lambda_{\chi 1} + N_2^A \lambda_{\chi 2}
~ ,
\end{align}
where these couplings are exactly the same ones as defined in Eq.~(\ref{HiggsCouplings}). By following this procedure for each of the \emph{du},  \emph{ud},  \emph{dd},  \emph{uu} models, we find:
\begin{align}
\lambda_{\chi h}^{du} &= \frac{1}{2} y^2 v \frac{ m_\chi \left(1+\cos{2 \beta} \cos{2 \theta} \right) +  M_D \sin{2 \beta} \sin{2 \theta} }{2 M_D^2  +4 M_S m_\chi - 6 m_\chi^2 + \frac{1}{2} y^2 v^2 \left( 1+ \cos{2 \beta} \cos{2 \theta} \right)}
\nonumber \\
\nonumber \\
\lambda_{\chi H}^{du} &= \frac{1}{2} y^2 v \frac{ m_\chi \sin{2 \beta} \cos{2 \theta} -  M_D \cos{2 \beta} \sin{2 \theta} }{2 M_D^2  +4 M_S m_\chi - 6 m_\chi^2 + \frac{1}{2} y^2 v^2 \left( 1+ \cos{2 \beta} \cos{2 \theta} \right)}
\nonumber \\
\nonumber \\
\lambda_{\chi A}^{du} &= \frac{1}{2} y^2 v  \frac{m_\chi \sin{2\beta} + M_D \sin{2\theta}}{2M_D^2 + 4 M_S m_\chi - 6 m_\chi^2 + \frac{1}{2}y^2v^2\left( 1+ \cos{2 \beta} \cos{2 \theta} \right) }
\label{duHiggs}
\end{align}

\begin{align}
\nonumber \\
\lambda_{\chi h}^{ud} &= y^2 v \frac{m_\chi \left( \sin^2{\beta} \cos^2{\theta} + \cos^2{\beta} \sin^2{\theta} \right) + \frac{1}{2} M_D \sin{2\beta} \sin{2\theta} }{2 M_D^2  +4 M_S m_\chi - 6 m_\chi^2 + \frac{1}{2} y^2 v^2 \left( 1- \cos{2 \beta} \cos{2 \theta} \right)}
\nonumber \\
\nonumber \\
\lambda_{\chi H}^{ud} &= - \frac{1}{2} y^2 v \frac{m_\chi \sin{2 \beta} \cos{2 \theta} + M_D \cos{2 \beta} \sin{2 \theta}}{2M_D^2 + 4 M_S m_\chi - 6 m_\chi^2 + \frac{1}{2}y^2 v^2 \left( 1- \cos{2 \beta}\cos{2 \theta} \right) }
\nonumber \\
\nonumber \\
\lambda_{\chi A}^{ud} &= -\frac{1}{2}y^2 v \frac{m_\chi  \sin{2\beta}+  M_D \sin{2\theta}}{2M_D^2 + 4M_S m_\chi - 6 m_\chi^2 + \frac{1}{2} y^2 v^2 \left( 1-\cos{2 \beta} \cos{2 \theta} \right)}
\label{udHiggs}
\end{align}

\begin{align}
\nonumber \\
\lambda_{\chi h}^{dd} &= y^2 v \cos^2{\beta} \frac{m_\chi + M_D \sin{2 \theta}}{2M_D^2 + 4 M_S m_\chi - 6m_\chi^2 + y^2v^2 \cos^2{\beta}}
\nonumber \\
\nonumber \\
\lambda_{\chi H}^{dd} &= \frac{1}{2} y^2 v \sin{2\beta} \frac{m_\chi + M_D \sin{2 \theta}}{2M_D^2 + 4 M_S m_\chi - 6m_\chi^2 + y^2v^2 \cos^2{\beta}}
\nonumber \\
\nonumber \\
\lambda_{\chi A}^{dd} &= \frac{1}{2}y^2 v \sin{2 \beta} \frac{m_\chi \cos{2 \theta} }{2M_D^2 + 4M_S m_\chi - 6 m_\chi^2 + y^2v^2 \cos^2{\beta}}
\label{ddHiggs}
\end{align}

\begin{align}
\nonumber \\
\lambda_{\chi h}^{uu} &= y^2 v \sin^2{\beta} \frac{m_\chi + M_D \sin{2 \theta}}{2M_D^2 + 4 M_S m_\chi - 6m_\chi^2 + y^2v^2 \sin^2{\beta}}
\nonumber \\
\nonumber \\
\lambda_{\chi H}^{uu} &= -\frac{1}{2} y^2 v \sin{2\beta} \frac{m_\chi + M_D \sin{2 \theta}}{2M_D^2 + 4 M_S m_\chi - 6m_\chi^2 + y^2v^2 \sin^2{\beta}}
\nonumber \\
\nonumber \\
\lambda_{\chi A}^{uu} &= -\frac{1}{2}y^2 v \sin{2 \beta} \frac{m_\chi \cos{2 \theta} }{2M_D^2 + 4M_S m_\chi - 6 m_\chi^2 + y^2v^2 \sin^2{\beta}}
~.
\label{uuHiggs}
\end{align}
The \emph{ud} model is related to the \emph{du} model by $\theta \to \pi/2 - \theta$ along with the pseudoscalar coupling, $\lambda_{\chi A}$, flipping sign. The \emph{uu} model, instead, is related to the \emph{dd} model by $\beta \to \beta + \pi/2$.

\twocolumngrid
\bibliography{monobUV}

\end{document}